\documentclass[journal]{IEEEtran}
\usepackage{footnote}
\makesavenoteenv{tabular}

\usepackage{amsfonts}
\usepackage{amsmath,amsfonts, amssymb}
\usepackage{color}
\usepackage{algorithm}
\usepackage{algorithmicx}
\usepackage{algpseudocode}
\usepackage{array}

\usepackage[caption=false,font=normalsize,labelfont=sf,textfont=sf]{subfig}
\usepackage{textcomp}
\usepackage{stfloats}
\usepackage{url}
\usepackage{verbatim}
\usepackage{graphicx}
\usepackage{cite}
\usepackage[hidelinks]{hyperref} 
\usepackage{cleveref} 
\usepackage{arydshln} 
\usepackage{tikz}
\usepackage{subfig}
\usepackage{pgfplots}
\usepackage{pgfplotstable}
\pgfplotsset{compat=1.17} 
\usetikzlibrary{pgfplots.statistics}
\newcommand{\blue}[1]{\textcolor{black}{#1}}
\definecolor{darkgreen}{rgb}{0.0, 0.5, 0.0} 
\newcommand{\green}[1]{
    {\color{black}#1}
}

\begin{document}

\title{Combining X-Vectors and Bayesian Batch Active Learning: Two-Stage Active Learning Pipeline for Speech Recognition}

\author{Ognjen~Kundacina,
        Vladimir~Vincan,
        Dragisa~Miskovic 
\thanks{
This work is funded by Serbian Ministry of Science, Technological Development and Innovation, through the Science and Technological Cooperation program Serbia-China, Research and development project No 00101957 2025 13440 003 000 620 021.

The authors are with The Institute for Artificial Intelligence Research and Development of Serbia, Novi Sad, 21000, Serbia (e-mail: ognjen.kundacina@ivi.ac.rs; vladimir.vincan@ivi.ac.rs; dragisa.miskovic@ivi.ac.rs).}}



\maketitle

\begin{abstract}
This paper introduces a novel two-stage active learning (AL) pipeline for automatic speech recognition (ASR), combining unsupervised and supervised AL methods. The first stage utilizes unsupervised AL by using x-vectors clustering for diverse sample selection from unlabeled speech data, thus establishing a robust initial dataset for the subsequent supervised AL. The second stage incorporates a supervised AL strategy, with a batch AL method specifically developed for ASR, aimed at selecting diverse and informative batches of samples. Here, sample diversity is also achieved using x-vectors clustering, while the most informative samples are identified using a Bayesian AL method tailored for ASR with an adaptation of Monte Carlo dropout to approximate Bayesian inference. This approach enables precise uncertainty estimation, thereby enhancing ASR model training with significantly reduced data requirements. Our method has shown superior performance compared to competing methods on homogeneous, heterogeneous, and OOD test sets, demonstrating that strategic sample selection and innovative Bayesian modeling can substantially optimize both labeling effort and data utilization in deep learning-based ASR applications.

\end{abstract}


\section{Introduction}
\label{sec_intro}

State-of-the-art transformer models for automatic speech recognition (ASR) require substantial volumes of labeled data. While there is an abundance of unlabeled speech recordings, high-quality labeled data is scarce, particularly in specialized domains or low-resource scenarios. Speech data labeling, however, is labor-intensive and time-consuming, potentially requiring more than eight hours to transcribe a single hour of speech recordings \cite{McMullin2023TranscriptionTime}, illustrating a significant bottleneck in the data preparation pipeline. Embracing a data-centric AI approach \cite{NEURIPS2023_DataCentricAI, DataCentricAI} by strategically choosing and labeling a diverse and informative subset of audio samples can result in a successfully trained and accurate ASR model, while reducing the labeling effort. Study \cite{DataSelection2007} reports that training the ASR model with just $17\%$ of carefully selected samples can attain the similar level of accuracy as training with the entire dataset, demonstrating the impact of data quality over quantity and enhancing the computational efficiency and data usage during the deep learning workflows.

Active learning (AL) is a technique that employs various aspects of machine learning models, such as outputs, gradients, and uncertainties, alongside the intrinsic properties of the unlabeled data samples, to select (query) the most informative or diverse points for labeling. AL minimizes the need for human involvement in data labeling, and is especially beneficial in datasets where duplicates, examples with little informational value, or noisy instances are common. Beyond failing to contribute to the model's training process, the presence of duplicate entries in the training dataset can negatively impact model performance \cite{lee-etal-2022-deduplicating}, underscoring the importance of implementing an intelligent data selection strategy through AL. Additionally, the importance of AL from the data management perspective for the enhancement of data preparation pipelines is highlighted in \cite{DataManagementForML2023}. Our work focuses on exploring AL methods designed for deep learning, known as deep active learning (DAL) \cite{2021SurveyDeepActiveLearning}, aiming to formulate effective AL algorithms for transformer-based models in ASR.

\textbf{Literature review}:
AL has been employed in ASR research as traditional ASR methodologies have advanced, utilizing methods based on discrete, symbolic representations of speech to estimate audio sample informativeness from statistical models \cite{AL_ASR_2005, YU2010GlobalEntropy}. Recently, deep learning-based models have become the preferred approach for ASR \cite{endToEndASRSurvey2024}, leveraging continuous, high-dimensional representations learned from large amounts of labeled data. As a result, there's a growing development of DAL techniques for deep learning-based ASR models, aimed at improving data quality and efficiency.

The most common application of DAL in ASR is found in supervised AL methods \cite{2021SurveyDeepActiveLearning}. These methods require an initial set of labeled data to train a deep learning model. The model then assesses the uncertainty of each unlabeled sample, selecting those with the highest uncertainty for further labeling. With these additional labeled samples, a new deep learning model instance is trained, which then more effectively identifies uncertain samples. This iterative process is usually repeated a predefined number of times. Supervised AL enables the model to improve its accuracy in areas of the data distribution where it initially performs poorly.

In the context of ASR, path probability-based AL methods evaluate the informativeness of unlabeled data samples by analyzing the decoded sequence probabilities of each token. These methods follow the principle that lower probabilities indicate higher informativeness or uncertainty, making such samples valuable for enhancing model training upon being labeled and added to the training set. Therefore, in this specific context, 'informativeness,' 'uncertainty,' and 'confidence' are often used interchangeably. Studies like \cite{Bang2020BoostingAL, drugman16_interspeech, Indian2018AL_ASR} apply these AL approaches in ASR, calculating the confidence levels of predictions based on path probabilities, or the posterior probabilities of tokens, derived from the softmax layer during sequence decoding.

The methods discussed above employ confidence scoring for AL, considering an unlabeled sample informative either due to its low path probability or high entropy across all labels, indicating uniform distribution of predictions. However, deep neural networks (DNNs) often exhibit overconfidence in predictions from the softmax layer \cite{Calibration_DNNs2017}, questioning the reliability of using path probabilities for uncertainty assessment of unlabeled data. A potential solution to this issue are gradient-based AL methods, which have also been applied for ASR \cite{jiaji2016Gradients1}. These methods evaluate informativeness by measuring the gradient norm of the model when processing unlabeled samples. The expected gradient length (EGL) metric estimates how much a model's parameters might change if the unlabeled sample was included in the training process. Following a similar path, the study in \cite{Yuan2019Gradients2} employs a distinct DNN to estimate the true gradient length, using EGL and entropy as inputs. The loss prediction method \cite{2021LossPrediction} operates similarly to gradient-based techniques in that it does not calculate uncertainty based on path probabilities, but provides an estimate of how much a model's parameters might change if the unlabeled sample were included in the training process. It involves joint training of the ASR model and a loss prediction DNN, where the uncertainty of an unlabeled sample is defined by the predicted loss function value.

To enhance the diversity of the methods previously mentioned, i-vectors \cite{i_vectors}, which are vector representations capturing speech audio sample variability, can be used. In \cite{malhotra19_i_vectors}, the least confidence method is expanded by incorporating an entropy measure related to clustered i-vectors. By clustering i-vectors into $K$ groups using the K-means algorithm, an entropy-based regularization term is added to the least confidence score. This entropy, associated with the distribution of selected samples across clusters, drives the selection of a diverse set of samples by encouraging sampling from all clusters. Implementing this strategy involves the tuning of an additional hyperparameter to balance the proposed diversity-based regularization term with the least confidence score.

Committee-based methods introduce an advanced approach by leveraging multiple models to assess the informativeness of unlabeled data samples, offering improvements over single-model confidence score methods discussed so far. The study \cite{LF_MMI_interspeech2018} integrates informativeness scores from both confidence-based and committee-based methodologies. In this setup, a committee comprising two DNNs, a recurrent neural network (RNN) and a time delayed neural network, trained under identical conditions, evaluates unseen samples. Their output transcriptions are compared to calculate the word matching error rate (WMER), which, when combined with the confidence score from the RNN, identifies the most informative unlabeled samples for selection. Further advancements are proposed in \cite{LMC-SMCA_2021_Committee1, 2019CommiteeDropout}, where the committee consists of the base model and models modified using dropout techniques. The discrepancy in transcriptions between these models quantifies uncertainty. Notably, \cite{2019CommiteeDropout} focuses on samples at the intersection of those chosen by the committee approach and those identified through the lowest confidence measures, to refine sample selection further.

All the methods discussed so far have been within the scope of supervised AL. Unsupervised AL, in contrast, does not require an initial set of labeled data.\footnote{In the literature, this concept is often referred to as \textit{cold-start active learning}. In our work, however, we will use the term unsupervised AL, as it has been previously adopted in the ASR domain \cite{unsupervised_AL_ASR}.} These methods may involve strategies like clustering or density estimation to identify valuable or representative samples from the unlabeled dataset. The study \cite{unsupervised_AL_ASR} presents an unsupervised AL method for ASR that employs acoustic features from a traditional ASR system and a KL-divergence-based density method to select a core subset of unlabeled samples that closely reflects the overall dataset's distribution. It suggests that an ASR model trained on a dataset chosen through unsupervised AL can perform similarly to the one trained using supervised AL, when the latter starts with a low-quality initial dataset. A recent work \cite{unsupervisedAL_interspeech2023} presents an approach that achieves unsupervised AL in a distinct setup, employing an ASR model trained on separate, unlabeled audio and text data, i.e., in an unsupervised manner \cite{NEURIPS2021_UnsupervisedASR}. This methodology diverges from the diversity-focused strategies typically associated with unsupervised AL, since it calculates an uncertainty metric for AL using the perplexity of phone sequences from the mentioned ASR model. This aspect of their approach shares similarities with the principles of supervised AL, while it leverages an ASR model trained in an unsupervised way.

In addition to AL, self-supervised learning has emerged as a popular approach for efficiently utilizing large volumes of unlabeled data in the ASR context \cite{hubert2021, selfSupervisedForASR2024}. In our work, we show that these two approaches could be combined for deep learning ASR models such as wav2vec 2.0 \cite{wav2vec2}, which are pre-trained in a self-supervised manner using unlabeled data to learn rich audio representations, and require subsequent fine-tuning on audio-transcription pairs. In other words, we will utilize a wav2vec 2.0 model pretrained without any ASR-specific training and fine-tune it with labeled data for the ASR task—data that has been effectively selected using AL techniques.

\begin{figure*}[htbp] 
\centering
\includegraphics[width=\textwidth, trim={0.5cm 7.3cm 1.4cm 0.55cm}, clip]{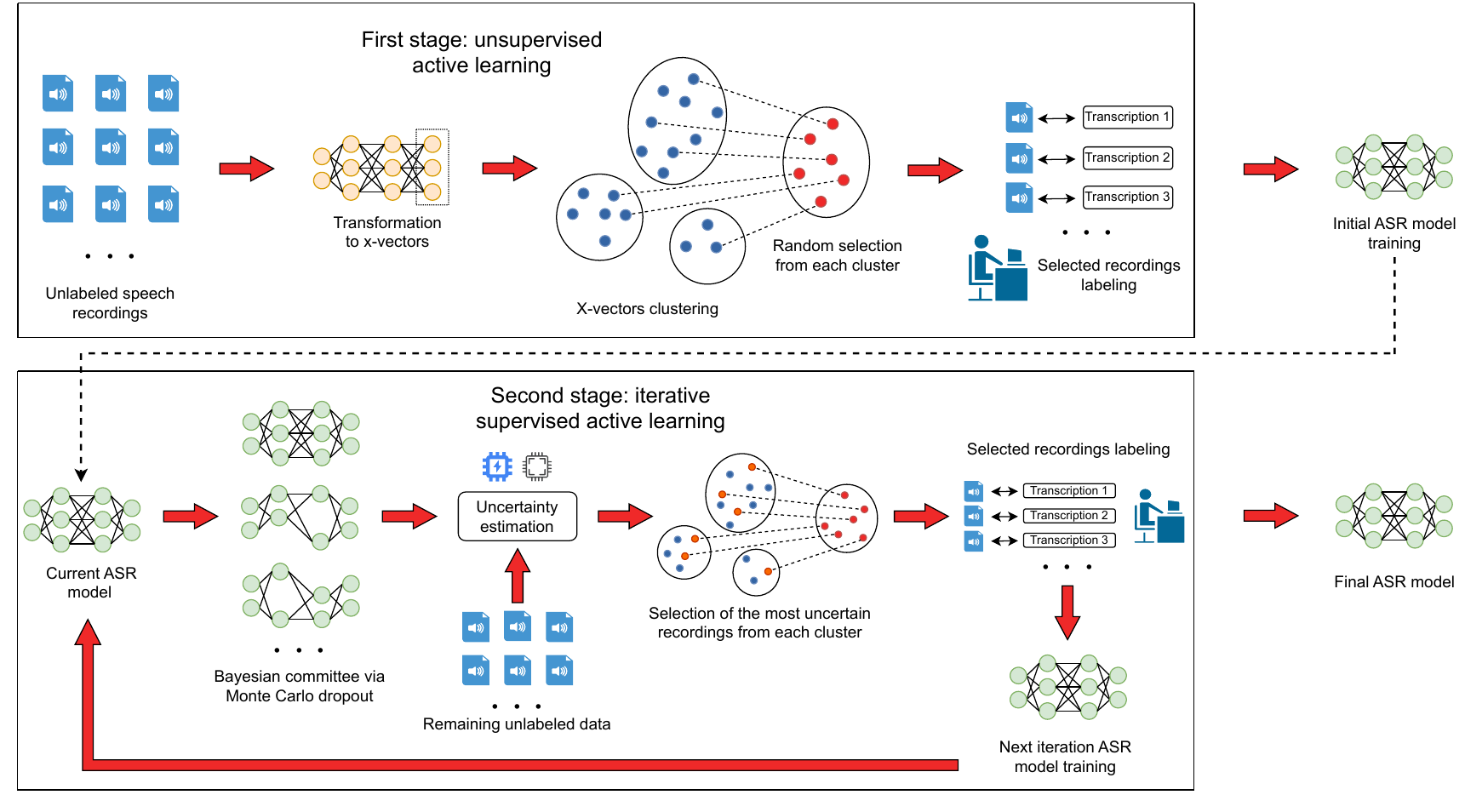}

\caption{The proposed two-stage active learning pipeline. The first, unsupervised active learning stage is based on x-vectors clustering. The second, supervised active learning stage combines x-vector-based batch active learning with Bayesian active learning via Monte Carlo dropout approximation.}
\label{fig:two_stage_pipeline}
\end{figure*}

\textbf{Contributions}:
Supervised AL algorithms typically initiate with a labeled dataset to train an initial deep learning model. This model then helps identify which subset of data would be the most beneficial for labeling next. In our research, we explore scenarios where AL starts with a completely unlabeled dataset. \blue{Currently, there are no studies applying a preparatory stage to efficiently select the initial labeled dataset for supervised AL in ASR.} We propose a two-stage AL process, presented on a high level in Fig. \ref{fig:two_stage_pipeline}, where the first stage involves using unsupervised AL to select the initial dataset for labeling. Our goal is to create an initial ASR model that is more effective than one trained on a similarly sized, but randomly chosen dataset. Subsequently, this initial dataset and the ASR model trained on it serve as inputs to the second stage of AL, which employs the supervised AL method.

For unsupervised AL, we employ x-vectors \cite{x_vectors}, extracted from a DNN trained for speaker classification tasks, offering advantages over the previously mentioned i-vectors. X-vectors provide a more nuanced representation of speech data, capturing deeper characteristics beneficial for distinguishing between different speakers and conditions. This makes them particularly effective for clustering purposes, allowing us to select diverse data samples in an unsupervised AL setting. Unlike the existing use of i-vector clusters in a supervised AL framework to enhance diversity \cite{malhotra19_i_vectors}, which necessitates tuning an additional hyperparameter to balance between uncertainty and selection from diverse clusters, the proposed approach employs direct sampling from x-vectors clusters to facilitate the selection of varied data samples. This approach simplifies the process, eliminating the need for additional hyperparameter tuning, while broadening the scope of data considered for an initial ASR model training. Moreover, our method ensures that each cluster is represented in the selection process, regardless of its size, guaranteeing that the ASR model is exposed to a broad range of data variations during the training process.

Following the initial ASR model training with a dataset selected using unsupervised AL, the second phase introduces an iterative supervised AL process. Each AL iteration leverages the current ASR model to identify a new subset of data for labeling using the proposed batch AL technique, tailored for ASR. Our adoption of batch AL, which queries a batch of multiple unlabeled samples in a single AL iteration, is motivated by its ability to select a diverse set of samples rather than just individually informative ones. This approach is particularly beneficial when faced with multiple similar uncertain samples, as it opts for a varied batch, enhancing the overall learning strategy. In contrast, non-batch AL methods might choose top-ranked samples based solely on confidence scores, potentially leading to a less diverse selection.

Our proposed batch AL approach for ASR uniquely integrates x-vectors clustering for sample diversity with uncertainty assessment through a Bayesian committee, specifically adjusted for ASR tasks. Theoretically, Bayesian neural networks \cite{BayesianNNbook} can calculate a model's uncertainty precisely, offering insights into which samples could most improve model performance. However, due to the high computational complexity of working with true posterior distributions in practice, we approximate the Bayesian neural network for ASR using Monte Carlo (MC) dropout \cite{pmlr-MC_dropout}. This approximation creates a distribution over transcriptions by employing a committee of models with diverse topologies.\footnote{It is important to note here that this is not a typical committee-based AL approach, since topologies of models are different for each new data sample.} These models are obtained by applying random dropout masks to simulate diverse predictive outcomes. Each model in the committee outputs a transcription for a given data sample, collectively forming a distribution. The variance within this transcription distribution is quantified by calculating the word error rate (WER) among the transcriptions, providing a measure of uncertainty for the given unlabeled data sample.

The contributions of this work can be summarized as follows:
\begin{itemize}
\item This work introduces, to our knowledge, the first two-stage AL pipeline for ASR that sequentially combines unsupervised and supervised AL, using unsupervised AL to establish a strategically selected initial dataset for enhancing the effectiveness of supervised AL afterward.
\item We employ unsupervised AL to sample from groups of unlabeled samples created through clustering of corresponding x-vectors \cite{x_vectors}, introducing a novel application of x-vectors in the context of AL for ASR. This approach shares some similarity with a method that employs i-vector clusters for regularization in a supervised AL setting \cite{malhotra19_i_vectors}; however, since our method is applied in an unsupervised AL context, it eliminates the need for an additional hyperparameter to balance regularization intensity. \green{As part of the unsupervised AL process, we perform disproportionate cluster sampling that favors underrepresented speaker groups, enforcing diversity and ensuring that all clusters are represented in the data selection.}
\item Our batch AL method, designed specifically for ASR and positioned within supervised AL, innovatively combines Bayesian AL with x-vectors clustering to ensure sample diversity and enable precise uncertainty estimation. \green{We achieve batch AL by clustering x-vectors and selecting a certain number of the most uncertain samples from each cluster. The number of samples selected from each cluster is determined using disproportionate cluster sampling, which prioritizes underrepresented speaker groups and ensures balanced data selection.} Importantly, the most computationally intensive part of our method, the uncertainty calculation, can be parallelized since uncertainties for each sample are computed independently.
\item For uncertainty calculation, we introduce a novel adaptation of Bayesian AL tailored for ASR. This adaptation includes approximating Bayesian inference using a Monte Carlo dropout committee and developing a method to calculate the WER-based variance over the transcription distribution provided by the committee, marking a unique approach in the field. \green{Compared to Bayesian active summarization \cite{GIDIOTIS2024BayesianActiveSummar}, which uses a similar uncertainty calculation for text by relying on pairwise comparisons of BLEU scores among committee members, our method uses a reference transcription without dropout, achieving linear computational complexity.}
\item The proposed approach is evaluated on a homogeneous test set designed to assess performance for groups of speakers underrepresented in the training data, and on a heterogeneous out-of-distribution (OOD) test set to test robustness, with our method consistently outperforming competing approaches in both scenarios. \green{Additionally, it is tested on a standard ASR evaluation benchmark, where its performance remains competitive with existing methods. While the first stage initially lags behind competing approaches, the second stage consistently achieves the best results in later iterations, demonstrating that the improvements observed on underrepresented and OOD test sets do not come at the cost of overall ASR performance.}
\end{itemize}

The paper is organized as follows: Section 2 provides the theoretical foundation for our approach, Section 3 details our methodology, Section 4 evaluates the effectiveness of our methods, and Section 5 summarizes our findings.

\section{Theoretical Preliminaries} \label{sec_theoretical}
In this section, we outline the theoretical background required for introducing our proposed approach. Specifically, we briefly discuss ASR utilizing the wav2vec 2.0 transformer neural network \cite{wav2vec2}, x-vectors \cite{x_vectors}, and the principles of DAL.

\subsection{Automatic Speech Recognition}

Modern ASR methods focus on the goal of transforming spoken language into text primarily using deep learning. The fundamental aim within this domain is to develop neural networks that can accurately interpret and transcribe human speech. In mathematical terms, an ASR system is represented as a trainable function \(f_{\theta}: \mathcal{X} \rightarrow \mathcal{Y}\), where \(\mathcal{X}\) is the set of all possible input audio samples and \(\mathcal{Y}\) is the set of all possible transcriptions, where \(\theta\) denotes trainable parameters. The training process involves adjusting the parameters \(\theta\) to minimize the difference between the predicted output and the actual transcription across a labeled training dataset \(D_{L} = \{(x_1, y_1), (x_2, y_2), \ldots, (x_N, y_N)\}\), where \(x_i \in \mathcal{X}\) denotes the $i^{th}$ audio input and \(y_i \in \mathcal{Y}\) represents its corresponding transcription.

Wav2vec 2.0, a transformer-based ASR model, processes raw audio waveforms directly \cite{wav2vec2}. Its architecture includes an encoder for extracting audio signal representations, which are then mapped to speech units by a subsequent linear layer. The model utilizes connectionist temporal classification (CTC) loss \cite{Graves2006CTC} for training, facilitating the alignment of audio sequences with text transcriptions without the need for explicit segmentation. While the integration of language models can enhance the transcription quality of wav2vec 2.0, they will not be considered in this work. Notably, wav2vec's design facilitates self-supervised pretraining by using segments of the audio input to predict adjacent segments, effectively learning the structure of audio data without labeled examples. In our work, we utilize wav2vec 2.0, pretrained in a self-supervised manner on unlabeled audio data, and subsequently fine-tune it using a labeled dataset selected through AL.

\subsection{X-Vectors}
X-vectors are embeddings used for speaker recognition, derived from DNNs trained to classify different speakers \cite{x_vectors}. The process involves a DNN that takes variable-length speech recordings and produces fixed-dimensional embeddings. This is achieved through a pooling layer that aggregates features across multiple frames early in the network. The architecture of the DNN includes several hidden layers after the pooling operation, where each layer is capable of representing the speaker's unique features as embeddings. The objective during training is to use these embeddings for speaker classification, employing a softmax layer. The softmax layer is discarded after training, since the focus lies on embeddings themselves, rather than the classification output. For training, the DNN utilizes a dataset with labeled speakers and applies multiclass cross-entropy loss to optimize the network's parameters. Data augmentation is employed to enhance the model's ability to generalize to new, unseen data, which is particularly relevant for our experiments that utilize datasets not previously encountered by the trained x-vectors.

Mathematically, the transformation performed by the DNN (excluding the softmax layer) on an audio data sample can be expressed as \(f_{\theta_x}: \mathcal{X} \rightarrow \mathbb{R}^d\), where \(d\) denotes the dimensionality of the output x-vector embeddings. This function \(f_{\theta_x}\), parameterized by the DNN's weights $\theta_x$, encapsulates the transformation of speech utterance into a compact, fixed-dimensional representation, serving as a fundamental component for selecting diverse samples in our AL approach.

\subsection{Deep Active Learning} \label{deep_AL}

Deep active learning optimizes the training of deep learning models by selectively labeling the most informative samples from an unlabeled dataset \cite{2021SurveyDeepActiveLearning}. Given a labeled dataset \(D_L = \{(x_i, y_i)\}_{i=1}^{|D_L|}\) where each \(x_i \in \mathcal{X}\) corresponds to an input sample with its label \(y_i \in \mathcal{Y}\), and an unlabeled dataset \(D_U = \{x_j\}_{j=1}^{|D_U|}\) where \(x_j \in \mathcal{X}\) lacks a label, the goal of DAL is to select a subset \(S \subseteq D_U\) for labeling that maximizes the model's performance improvement, by exploiting $D_L$ if possible.

In supervised AL, the model is initially trained on the existing labeled data \(D_L\) and then iteratively updated by incorporating samples from \(S\) that, once labeled and added to \(D_L\), are expected to most significantly reduce the model’s error. More formally, supervised AL defines an acquisition function, depending on the specific AL method or uncertainty metric, to select the samples with the largest contribution to the model's performance in each AL iteration. This iterative process continues until achieving a predefined performance level or exhausting the labeling budget. On the other hand, unsupervised AL does not rely on \(D_L\) for initial training, but instead focuses on the structure or distribution of \(D_U\) to select data samples for labeling. Techniques such as clustering or density estimation might be used to identify representative samples.

In supervised AL, the \textbf{uncertainty-based} methods selects samples from the unlabeled dataset $D_U$ for which the model's current predictions are the least confident. This strategy aims to identify the samples that, if labeled and added to the training set $D_L$, would most improve the model's performance by reducing its overall uncertainty. One common uncertainty-based acquisition function for classification tasks is entropy \cite{Active_learning_survey}, which for a given sample \(x \in D_U\) is defined as:
\begin{equation}
H(x) = - \sum_{c=1}^{C} P(y_c|x) \log P(y_c|x),
\end{equation}
where \(P(y_c|x)\) is the model's predicted probability that the sample \(x\) belongs to a class \(c\), and \(C\) is the total number of classes. The uncertainty-based query strategy selects a subset \(S\) from \(D_U\) consisting of a predefined number of samples with the highest entropy values for labeling.

\textbf{Committee-based} methods in supervised AL use a group of diverse models, named a committee, to make collective decisions on which samples to label next \cite{DAGAN1995Committee}, mitigating individual model biases that may occur in AL methods that rely solely on a single model's prediction probabilities. This approach is based on the idea that different models may disagree on the predictions for the same unlabeled data point, and these disagreements can highlight areas where acquiring labels would most benefit the learning process. Given a committee of models \(\{f_{\theta_1}, f_{\theta_2}, \ldots, f_{\theta_M}\}\), where each \(f_{\theta_m}\) represents a model with parameters \(\theta_m\) and \(M\) is the number of models in the committee, the uncertainty of the committee for a sample \(x \in D_U\) can be quantified in several ways. One common acquisition function for committee-based methods is the vote entropy \cite{DAGAN1995Committee}, defined for a classification task with \(C\) classes as:

\begin{equation}
V(x) = -\sum_{c=1}^{C} P_{\text{vote}}(y_c|x) \log P_{\text{vote}}(y_c|x)
\end{equation}
where \(P_{\text{vote}}(y_c|x)\) is the proportion of models in the committee that predict \(x\) as belonging to class \(c\). This vote entropy increases with the level of disagreement among the committee members about the class of \(x\). The committee-based query strategy then selects a subset \(S\) from \(D_U\), consisting of samples with the highest vote entropy values for labeling.

In \textbf{Bayesian active learning}, weights within the DNN model are represented as probability distributions, rather than deterministic estimates. This approach allows for the estimation of uncertainty in data samples by examining the variability in the model's outputs, as Bayesian DNNs generate output distributions. A higher variability in these outputs for unlabeled data samples indicates greater uncertainty, suggesting the prioritization of samples for labeling \cite{Houlsby2011BayesianAL}. The uncertainty is inferred from the posterior distribution of the model parameters, \(P(\theta | D_L)\), which is updated upon observing new data \(D_L\) in the following way:
\begin{equation}
    P(\theta | D_L) = \frac{P(D_L | \theta) P(\theta)}{P(D_L)},
\end{equation}
where \(\theta\) denotes the parameters of the model. The process updates the posterior distribution \(P(\theta | D_L)\) by combining the initial (often Gaussian) prior beliefs \(P(\theta)\) regarding the parameters with the likelihood of the data given the parameters \(P(D_L | \theta)\). Additionally, it involves the marginal likelihood of observing the data \(P(D_L)\), which is integrated over all possible parameter values \(\theta\), and is computationally demanding to perform in practice.

In Bayesian AL, the predictive distribution can be used for quantifying the uncertainty associated with model predictions, providing a probabilistic basis for making decisions about which data samples to label next. The exact predictive distribution for output \( y \), given an input \( x \) and the observed labeled dataset \( D_L \), is calculated by integrating the likelihood \( P(y|x, \theta) \) of observing \( y \) given the input \( x \) and model parameters \(\theta\), with the posterior distribution \( P(\theta|D_L) \) over all possible values of  parameters:
\begin{equation} \label{eq-predictive-distribution}
P(y|x, D_L) = \int P(y|x, \theta) P(\theta|D_L) d\theta.
\end{equation}

Accurately computing the true posterior \(P(\theta | D_L)\) and the predictive distribution \(P(y|x, D_L)\) in Bayesian neural networks can be computationally challenging, due to the need for integration across a high-dimensional parameter space. To overcome this, we can employ stochastic regularization techniques such as the Monte Carlo dropout as an approximation \cite{pmlr-MC_dropout}. This method uses dropout layers from the standard DNN architecture, with the assumption that the DNN is trained using dropout, to perform stochastic forward passes during inference. By performing multiple forward passes and aggregating the results, we simulate sampling from the posterior distribution of model parameters. This approximation, while maintaining the original DNN's structure unchanged, effectively captures the uncertainty by leveraging a randomly generated committee of models.

For regression problems, the uncertainty measure for a data sample is typically the variance across \(T\) predicted outputs, which is a direct approximation of the variance of the predictive distribution \(P(y|x, D_L)\) defined in \eqref{eq-predictive-distribution}, where \(T\) denotes the number of MC dropout forward passes. For classification problems, there are multiple ways to quantify output variability, using which various acquisition functions can be defined, such as the variation ratio, predictive entropy, or mutual information between predictions and the model posterior \cite{gal2016uncertainty}. For example, the variation ratio for a sample \( x \), which measures the data dispersion around the mode of distribution \( c^* \) obtained using \(T\) forward stochastic passes, is defined as: 
\begin{equation}
V_{\text{ratio}}(x) = 1 - \frac{1}{T} \sum_{t=1}^{T} \mathbb{I}[y_t = c^*].
\end{equation}
Here, $\mathbb{I}[\cdot]$ is the indicator function, and \( y_t \) is the predicted class for the \( t^{th} \) forward pass. As \( T \) grows, the expression \( \frac{1}{T} \sum_{t=1}^{T} \mathbb{I}[y_t = c^*] \) begins to approximate \( P(y = c^* | x, D_L) \), which is the maximum predicted probability for class \( c \) given the input \( x \) and the labeled data \( D_L \) that the DNN model was trained on, making it a valuable measure for a model's uncertainty in the context of AL.

\textbf{Batch active learning}, referenced also as batch-mode active learning in some studies, queries multiple unlabeled samples in one AL iteration \cite{NEURIPS2021_BatchAL_At_scale}, unlike traditional supervised AL methods that query one sample at a time. Given the substantial computational costs associated with retraining large and complex deep learning models, it is impractical to retrain after adding just a single labeled sample to the training set, particularly since such an addition typically results in insignificant improvements. To overcome this issue, traditional AL methods can be adjusted to select a batch of the most uncertain samples in every AL iteration. However, the samples are selected based on their individual uncertainties, without considering which combination would contribute to the training process the most. The main goal of batch AL is to select a predefined number \(b\) of samples that are both diverse and informative from the unlabeled dataset in each AL iteration. This approach is beneficial when the model encounters multiple similar samples with high uncertainty, avoiding the selection of the top uncertain $b$ samples only. Focusing both on informativeness and diversity reduces the chance of choosing duplicate samples, which could hinder learning progress, as indicated in \cite{lee-etal-2022-deduplicating}. Additionally, avoiding duplicates is beneficial because it prevents the waste of training and labeling resources on redundant information and ensures a broader coverage of samples in a single batch. 

Formally, in batch AL, a batch of data samples \( \{x^*_{1}, \ldots, x^*_{b}\} \) is selected from the unlabeled dataset \( D_U \) based on a generalized acquisition function \( A \). This function evaluates the candidate batches according to criteria that could include uncertainty, diversity, clustering outcomes, or their combination. The selection process can be formally described as:
\begin{equation} \label{eq_general_a}
\{x^*_{1}, \ldots, x^*_{b}\} = \underset{\{x_1, \ldots, x_b\} \subseteq D_U}{\mathrm{argmax}} \ A(\{x_1, \ldots, x_b\}, \phi),
\end{equation}
where parameters $\phi$ can include the current model parameters $\theta$ trained on the current labeled dataset $D_L$, as well as any additional metrics or heuristics relevant to the batch selection strategy.

\section{Proposed Approach} \label{sec_proposed}
In this section, we introduce our proposed approach, which includes the first stage of the AL pipeline using unsupervised AL via x-vectors, followed by the second stage that employs a Bayesian batch AL method, specifically tailored for ASR. The proposed pipeline starts with unsupervised AL to pick an initial dataset, setting the stage for more effective supervised AL. This chosen dataset and the ASR model trained on it initiate the second AL stage, where supervised AL is applied to select new, informative data samples for labeling. These labeled samples are then added to the training set, allowing for the retraining of the ASR model. This retrained model is then used to select additional data for labeling in the new AL iteration.

\subsection{First Stage: Unsupervised Active Learning} \label{first_stage}

The first AL stage begins with an unlabeled dataset \(D_U = \{x_i\}_{i=1}^{|D_U|}\). An x-vector DNN \(f_{\theta_x}\) \cite{x_vectors}, generates a corresponding set of x-vectors, producing one vector \(\mathbf{x}_i \in \mathbb{R}^d\) for each audio recording \(x_i\), formalized as:
\begin{equation} \label{eq_create_x_vectors}
\{\mathbf{x}_i\}_{i=1}^{|D_U|} = f_{\theta_x}(D_U).
\end{equation}

These x-vectors are then clustered using the DBSCAN density-based spatial clustering algorithm \cite{DBSCAN2017}, chosen for its ability to identify clusters of a configurable minimal size, thereby avoiding overly small clusters. DBSCAN identifies core samples for high density areas and extends clusters based on them. It groups together points that are nearby, while identifying and labeling as outliers those located in low-density areas. This is beneficial for isolating groups of audio samples that share common characteristics, such as being from the same or similar speakers, or similar acoustic conditions. Unlike approaches such as \cite{malhotra19_i_vectors} that employ K-means for clustering i-vectors, to the best of our knowledge, DBSCAN has not been applied to speech embeddings in the context of AL. DBSCAN's advantage lies in its ability to discover clusters of arbitrary shape without defining the desired number of clusters, its robustness to noise and outliers, and its insensitivity to the order in which data points are processed. This grouping of audio samples is performed based on the spatial relationships within the high-dimensional vector space generated by x-vector DNN \(f_{\theta_x}\). The result of DBSCAN clustering can be represented as a set of clusters \(\{C_k\}_{k=1}^{K}\), where each cluster \(C_k\) contains x-vectors that are more similar to each other than to those in other clusters.

In subsequent steps of the first stage of AL, we enhance diversity among the data samples selected for labeling by sampling from clusters of corresponding x-vectors. To select an initial set of samples for labeling \(S^0 \subseteq D_U\), we employ disproportionate cluster sampling, slightly favoring smaller clusters to ensure diversity. Here, the superscript $0$ indicates the iteration index within the entire AL process. This approach can be formalized as:
\begin{equation} \label{eq_samples_per_cluster}
|S^0_k| = \left\lceil \alpha_k(|C_k|) \cdot \frac{|C_k|}{\sum_{i=1}^{K}|C_i|} \cdot |S^0| \right\rceil,
\end{equation}
where \(|S^0_k|\) is the number of samples selected from a cluster \(C_k\) uniformly at random, and \(\alpha_k\) is an affine function of cluster size $|C_k|$:
\begin{equation} \label{eq_cluster_function}
\alpha_k(|C_k|) = \beta - \gamma \cdot \frac{|C_k|}{\sum_{i=1}^{K} |C_i|},
\end{equation}
This function introduces a variable slope to favor smaller clusters by assigning larger values of \(\alpha_k\) to them, and vice versa for larger clusters. Here, \(\beta\) and \(\gamma\) are parameters that adjust the extent to which cluster size influences sample selection. Finally, the set of samples selected for labeling is equal to the union of samples selected from each cluster: $S^0 = \bigcup_{k=1}^{K} S^0_k$, while the remaining unlabeled samples are given by the following formula:
\begin{equation} \label{eq_remaining_samples}
D_U^0 = D_U \setminus S^0.
\end{equation}
Following the selection, samples from \(S^0\) are labeled to form the initial labeled dataset \(D_L^0\), which, along with the ASR model \(f_{\theta^0}\) trained on it, serves as input for the following supervised AL stage. The pseudocode encapsulating the entire first stage of AL is presented in Algorithm~\ref{first_stage_alg}.

\begin{algorithm}[!b]
\caption{First stage: unsupervised AL}
\label{first_stage_alg}
\begin{algorithmic}[1]
        \State Input: unlabeled dataset \(D_U = \{x_i\}_{i=1}^{|D_U|}\)
        \For {$i=1,2,\ldots,|D_U|$}
            \State Calculate x-vector $\mathbf{x}_i$ for each $x_i$ using \eqref{eq_create_x_vectors} 
        \EndFor
        \State Cluster the set of x-vectors $\{\mathbf{x}_i\}_{i=1}^{|D_U|}$ using DBSCAN
        \State For each cluster $C_k$ from the resulting set \(\{C_k\}_{k=1}^{K}\) determine \(|S^0_k|\) using \eqref{eq_samples_per_cluster} and \eqref{eq_cluster_function}
        \State Form $S^0_k$ by selecting \(|S^0_k|\) samples from each cluster $C_k$ randomly
        \State $S^0 = \bigcup_{k=1}^{K} S^0_k$
        \State Label selected samples $S^0$ and add them to $D_L^0$
        \State Train an initial ASR model $f_{\theta^{0}}$ using $D_L^0$
        \State Update the unlabeled dataset $D_U^0$ using \eqref{eq_remaining_samples}
\end{algorithmic}
\end{algorithm}

We designed the unsupervised AL stage using x-vectors clustering for initial data selection, and we reuse these x-vectors clusters to diversify sample selection during the supervised AL phase. Additionally, this clustering-based unsupervised AL method represents an improvement over the supervised AL approach based on i-vector clustering \cite{malhotra19_i_vectors}, which requires tuning an extra hyperparameter $\lambda$ to balance between diversity and uncertainty. In contrast, our approach, applicable in both unsupervised and supervised AL stages, achieves sample diversity without requiring adjustments to this specific hyperparameter.

\subsection{Second Stage: Supervised Active Learning}

The second stage of AL is an iterative supervised AL approach that integrates concepts from both batch AL and Bayesian AL, tailored specifically for ASR applications. In this approach, we enhance sample informativeness by selecting samples on which the Bayesian committee shows the highest disagreement, and we promote sample diversity by choosing the most informative samples from each x-vectors cluster \(C_k\) defined in Section \ref{first_stage}.

\textbf{Bayesian AL for ASR:} The direct application of Bayesian AL described in Section \ref{deep_AL} to ASR presents challenges due to the sequential nature of speech. ASR differs from standard classification tasks due to its focus on making sequential predictions, where each predicted phoneme in a transcription is effectively a separate classification event. Since metrics designed for fixed-length predictions, such as the variation ratio, are not suitable for the sequence-to-sequence problems, we will employ WER metric to align our Bayesian AL method with the ASR problem.

As outlined in Section \ref{deep_AL}, Monte Carlo dropout serves as an approximation of the exact integration of the model posterior over the parameter space. It can also be interpreted as a Bayesian committee of models, where random dropout during multiple forward passes yields a diversity of model topologies. This ensemble of varied models outputs a range of possible transcriptions for a given input. For each audio sample \(x_i\) in the unlabeled dataset, we conduct a series of \(T\) stochastic forward passes using the ASR model \(f_{\theta^{h-1}}\) as a base, derived from the previous supervised AL iteration. The index \(h\) ranges from 1 to \(H\), with \(H\) denoting the total number of supervised AL iterations planned within the second stage of AL. Each stochastic forward pass, indexed by \(t\), applies a unique MC dropout mask, producing a transcription \(y_i^{t,h}\) as follows:

\begin{equation} \label{eq_yt}
y_i^{t,h} = f_{\theta_t^{h-1}}(x_i) \quad \text{for } t \in \{1, \ldots, T\},
\end{equation}
where \(f_{\theta_t^{h-1}}\) indicates the ASR model altered with MC dropout mask during the \(t^{th}\) forward pass. Additionally, a reference transcription \(y_i^{r,h}\) from the ASR model without dropout to serve as a stable baseline is given by:

\begin{equation} \label{eq_yr}
y_i^{r,h} = f_{\theta^{h-1}}(x_i).
\end{equation}

We then calculate the WER for each transcription \(y_i^{t,h}\) against the reference \(y_i^{r,h}\):

\begin{equation}  \label{eq_wer}
WER(y_i^{t,h}, y_i^{r,h}) = \frac{S_i^{r,t,h} + D_i^{r,t,h} + I_i^{r,t,h}}{N_i^{r,h}},
\end{equation}
where \(S_i^{r,t,h}\) is the number of substitutions, \(D_i^{r,t,h}\) is the number of deletions, \(I_i^{r,t,h}\) is the number of insertions required to change the transcription $y_i^{t,h}$ into the reference \(y_i^{r,h}\), and \(N_i^{r,h}\) is the total number of words in the reference transcription. This calculation yields the individual WER for each pair of transcriptions, which we average across all \(T\) stochastic transcriptions to measure the overall audio sample uncertainty $U^h(x_i)$, which we use for guiding the selection of samples for labeling in the subsequent iteration of the supervised AL process:

\begin{equation} \label{eq_uncertainty}
U^h(x_i) = \frac{1}{T} \sum_{t=1}^{T} WER(y_i^{t,h}, y_i^{r,h}).
\end{equation}

This approach was inspired by the Bayesian active summarization strategy in \cite{GIDIOTIS2024BayesianActiveSummar}, which uses a pairwise bilingual evaluation understudy (BLEU) score variance among summary candidates gathered using $T$ stochastic forward passes as an uncertainty metric. However, our method contrasts in its computational efficiency, offering a complexity of \(\mathcal{O}(T)\) for each input sample compared to the \(\mathcal{O}(T^2)\) complexity from the pairwise BLEU score calculations. While the BLEU score is used for evaluating the fluency and adequacy of language translations, we choose WER as a more suitable metric for ASR, as it directly measures transcription accuracy at the word level, aligning with the precision requirements of ASR tasks.

\textbf{Batch AL for ASR:} In the proposed supervised AL framework tailored for ASR, batch diversity is achieved by using x-vectors clusters \( \{C_k\}_{k=1}^{K} \) obtained in the initial unsupervised AL stage, while the sample informativeness is ensured through the uncertainty metric \( U(x_i) \) as formulated in \eqref{eq_uncertainty}. The number of samples \( |S^h_k| \) to be chosen from each cluster $k$ in iteration $h$ is calculated by disproportionate cluster sampling, which favors smaller clusters slightly more, as described by Equations \eqref{eq_samples_per_cluster} and \eqref{eq_cluster_function}. For each cluster \( C_k \), we select a subset of samples \( S^h_k \) with the highest uncertainty values:
\begin{equation} \label{eq_top_uncertain}
S^h_k = \underset{\{x_1, \ldots, x_{|S^h_k|}\} \subseteq D_U}{\mathrm{argmax}} \ U^h(x_i).
\end{equation}
By collecting the top-ranked samples from all clusters, we form the batch \( S^h \) targeted for labeling in the current iteration of supervised AL.
\begin{equation}  \label{eq_top_uncertain_union}
S^h = \bigcup_{k=1}^{K} S^h_k = \bigcup_{k=1}^{K} \underset{\{x_1, \ldots, x_{|S^h_k|}\} \subseteq D_U}{\mathrm{argmax}} \ U^h(x_i).
\end{equation}
Upon labeling selected samples $x_i \in S^h_k$ with the corresponding transcriptions $y_i$, datasets of labeled and unlabeled samples are updated in the following way:
\begin{subequations} \label{eq_update_datasets}
  \begin{gather}
    D_L^h = D_L^{h-1} \cup \{(x_i, y_i) \mid x_i \in S^h \}, \label{eq_update_datasets_a}\\
    D_U^h = D_U^{h-1} \setminus S^h. \label{eq_update_datasets_b}
  \end{gather}
\end{subequations}

The pseudocode of the presented iterative procedure of the second stage of AL is given in Algorithm~\ref{second_stage_alg}. This procedure aligns with the general concept of batch AL as previously defined, where the acquisition function \( A \) given in \eqref{eq_general_a} balances the considerations of uncertainty and diversity within and across the clusters.

\begin{algorithm}[!b]
\caption{Second stage: supervised AL}
\label{second_stage_alg}
\begin{algorithmic}[1]
        \State Input: remaining unlabeled data \(D_U^0\), initial labeled data \(D_L^0\) and the trained ASR model $f_{\theta^{0}}$ 

        \For {$h=1,2,\ldots,H$}
            \For {$i=1,2,\ldots,|D_U^{h-1}|$}
                \State Load unlabeled audio sample $x_i$
                \State Generate Bayesian committee $\{f_{\theta_t^{h-1}}\}_{t=1}^{T}$ using 
                \Statex \hspace{\algorithmicindent}\hspace{\algorithmicindent} the base model \(f_{\theta^{h-1}}\) and MC dropout
                \State Generate Bayesian committee transcriptions $y_i^{t,h}$
                \Statex \hspace{\algorithmicindent}\hspace{\algorithmicindent} using \eqref{eq_yt}
                \State Generate reference transcription  $y_i^{r,h}$ using \eqref{eq_yr}
                \State Calculate $WER(y_i^{t,h}, y_i^{r,h})$ using \eqref{eq_wer}
                \State Calculate uncertainty $U^h(x_i)$ using \eqref{eq_uncertainty}
            \EndFor 
            \State For each cluster in \(\{C_k\}_{k=1}^{K}\) determine \(|S^h_k|\) 
            \Statex \hspace{\algorithmicindent} using \eqref{eq_samples_per_cluster} and \eqref{eq_cluster_function}

            \State Form $S^h_k$ by selecting top \(|S^0_k|\) uncertain samples 
            \Statex \hspace{\algorithmicindent} from each cluster $C_k$ using \eqref{eq_top_uncertain}
            \State Form the batch $S^h = \bigcup_{k=1}^{K} S^h_k$, as in \eqref{eq_top_uncertain_union}
            \State Label selected samples $S^h$
            \State Update datasets of labeled and unlabeled samples, 
            \Statex \hspace{\algorithmicindent} $D_L^h$ and $D_U^h$, using  \eqref{eq_update_datasets_a} and \eqref{eq_update_datasets_b}
            \State Train the ASR model \(f_{\theta^{h}}\) using $D_L^h$
            
        \EndFor
\end{algorithmic}
\end{algorithm}

\textbf{Performance optimization for large clusters:}
Although the proposed method for calculating uncertainty has linear computational complexity with respect to the number of samples in a cluster, execution time can still be substantial for clusters with a large number of samples. To address this, we propose two simple strategies to decrease execution time in such instances:
\begin{itemize}
    \item The most computationally demanding aspect of our approach, i.e. calculating uncertainty using a committee, can be efficiently parallelized, since the predictions of the committee members can be calculated independently. This stands in contrast to common batch AL methods, such as BatchBALD \cite{NEURIPS2019_BatchBALD}, where the mutual information between the model's predictions on a batch of samples and the model parameters must be calculated jointly. This inherently sequential process prevents  efficient parallelization, leading to increased computational demands as batch and cluster sizes grow.
    \item The research paper \cite{Atighehchian2020BAAL} demonstrates that for large unlabeled datasets, the performance of AL remains unaffected when it is applied to only $25\%$ of the unlabeled data, thereby achieving a considerable boost in efficiency. Our method can leverage this to significantly enhance efficiency by applying it to large unlabeled data clusters when necessary.
\end{itemize}

\section{Results and Discussion} \label{sec_results}
In this section, we first describe the datasets used, and outline the training hyperparameters employed for the wav2vec 2.0 ASR model. We proceed by presenting the results of the first, unsupervised AL stage, with the addition of discussing the advantages of using x-vectors over i-vectors. Next, we present the results of the second, supervised AL stage, additionally including the discussion on the quality of the proposed Bayesian-based uncertainty calculation. Finally, we review the results of the proposed approach on the OOD test set,\green{ as well as on a standard ASR train-test split}.

The English language subsets of the Common Voice 1.0 \cite{common-voice-2020-common} and the LibriSpeech \cite{LibriSpeech2015} ASR datasets were used for conducting experiments in this paper. The Common Voice english-train subset has a total duration of around 15 hours of annotated data, while the LibriSpeech train-clean-100 has around 100 hours of annotated data. The distributions of the datasets are very different, as can be seen in \Cref{table:datasets}. Notably, the Common Voice dataset is more heterogeneous, containing almost five times as many speakers while featuring more than twice fewer recordings compared to LibriSpeech. Within Common Voice, there are 426 speakers featured in singular audio files, whereas in LibriSpeech each speaker contributes in multiple audio files, the lowest of which being 26 files. In Common Voice, \blue{a single speaker occurs in 687 audio files with a total duration of 50.8 minutes}, whereas in LibriSpeech, this figure is 166 \blue{audio files with a total duration of 35.1 minutes.} The average duration of audio files is 2.85 times longer in LibriSpeech than in Common Voice.

When evaluating AL methods using heterogeneous test sets, performance can often appear similar across different strategies, largely due to the high variability inherent in these test sets. This variability can sometimes mask differences in method effectiveness, and even random sampling may yield comparably good performance, as demonstrated by \cite{Munjal2022Cvpr_random_sampling}. To primarily assess the real-world efficacy of the proposed AL pipeline, we have created a more homogeneous test set containing only LibriSpeech data, hereafter referred to as the \textit{primary test set}. Another goal of this test set is to assess the capability of the proposed AL approach to select a diverse data samples, particularly focusing on those that are underrepresented in the unlabeled dataset. This simulates a scenario that involves a specific class of speakers, representing potential users of the ASR system being developed, who might differ from the typical user base, aiming to ensure robust performance for them despite their limited presence. \green{In many real-world ASR applications, the performance of models varies significantly across different speaker groups, particularly for those with less common accents or dialects \cite{DiChristofano2024, tadimeti-2022, Graham2024, Allison2020}. These speaker groups often form smaller clusters in the feature space, making them less likely to be well-represented in training data. The primary test set aims to reflect this challenge by isolating speakers whose x-vector clusters do not overlap with the dominant speaker groups. While this setup is not a direct simulation of accent-based representation, it follows a similar principle where a subset of speakers with unique speech characteristics is underrepresented in the unlabeled dataset.}

To implement this test setup, we \green{identified all speakers in the LibriSpeech dataset whose x-vector clusters do not overlap with each other or with dominant speaker groups. Since ten speakers met this criterion, we selected all of them to ensure full representation of underrepresented speaker groups in our evaluation.} \green{For each selected speaker, we randomly selected $30\%$ of their speech samples for the primary test set, while the remaining $70\%$ were included in the unlabeled dataset.}\blue{\footnote{\blue{Although it is not common practice when evaluating ASR models to have the same speakers in both the training and test sets, this setup is reasonable in our case, as the primary test set consists exclusively of samples that are absent from the training set.}}} \green{This proportion ensures that enough data remains in the unlabeled pool for AL while providing a sufficiently large number of test samples to evaluate performance on underrepresented speakers. The unlabeled} dataset also includes \green{the train split of} the heterogeneous Common Voice dataset, which is larger by a factor of 20, effectively simulating the previously described scenario involving underrepresented speakers. \green{In total, the primary test set consists of 315 audio samples, while the unlabeled dataset contains 12868 samples.} While we expect that our AL approach will mostly select samples from the more represented unlabeled Common Voice data, we aim to assess whether it promotes diversity by suggesting informative training data samples from these small speaker groups and achieves strong performance on the corresponding test samples.

Additionally, we also test our approach using the VoxPopuli dataset \cite{wang2021voxpopuli} to evaluate its performance on both OOD and more heterogeneous data. In contrast to the LibriSpeech and Common Voice datasets, which consist of speakers reading aloud, the VoxPopuli dataset includes utterances from plenary sessions of the European Parliament, accompanied by aligned transcriptions. The content of these utterances, as well as the language specific to the European Parliament domain, contributes to a distribution shift, ideal for testing in an OOD setting. The OOD test set comprises 1842 audio files from 286 distinct speakers, making it the most heterogeneous dataset with roughly twice fewer audio files per speaker than Common Voice and nearly 20 times fewer than LibriSpeech, making it well-suited for evaluating performance on heterogeneous data. The maximum duration of each utterance is 20 seconds, with an average duration of 9.63 seconds, situated between the averages of the Common Voice and LibriSpeech datasets, while the summed duration over the whole dataset is about five hours.

\green{Finally, we also evaluate the proposed approach on a standard Common Voice 1.0 train-test split, representing a typical ASR benchmark. This evaluation assesses whether our method performs comparably to competing approaches on a widely used dataset, ensuring that the improvements observed on underrepresented and OOD test sets do not come at the cost of overall ASR performance. For this experiment, the unlabeled dataset, used for selecting data iteratively to train ASR models, consists of 12135 audio samples from the Common Voice train split, while the test set contains 7015 samples from the Common Voice test split.}

\begin{table}[h!]
\begin{center}
\caption{Comparison Between Datasets}
\label{table:datasets}
\begin{tabular}{ |c||c|c| } 
 \hline
 & Common Voice & LibriSpeech \\
 \hline
 duration [$s$] & 53832.70 & 362127.17 \\
 \# audio files & 12135 & 28539 \\
 \# speakers & 1057 & 251 \\
 \hdashline
 duration mean [$s$] & 4.44 & 12.69 \\ 
 duration variance [$s^2$] & 1.79 & 12.78 \\ 
 duration skewness [$s^3$] & 1.11 & -1.39 \\ 
 duration kurtosis [$s^4$] & 3.17 & 0.98 \\ 
 \hdashline
 speaker occurrence mean & 11.48 & 113.70 \\ 
 speaker occurrence variance & 2052.63 & 231.39 \\ 
 speaker occurrence skewness & 10.37 & -0.88 \\ 
 speaker occurrence kurtosis & 129.07 & 6.44 \\ 
 \hline
\end{tabular}

\end{center}
\end{table}

For our experiments, we utilized wav2vec 2.0 XLS-R model \cite{Babu2021XLSRSC}, which consists of 300 million trainable parameters and was pretrained on 436 thousand hours of unlabeled speech data. To ensure uniformity across all ASR training experiments, the same conditions and hyperparameters were applied, with key ones presented in Table \ref{table_hyperparams}. Additionally, in all the experiments we used slope parameters \(\beta = 0.095\) and \(\gamma = 0.0553\) in \eqref{eq_cluster_function}.\footnote{The code for reproducing the experiments is available at \href{https://github.com/VladimirVincan/active-learning-asr}{https://github.com/VladimirVincan/active-learning-asr}.} Finally, during the implementation of experiments, we utilized the MC dropout implementation from the Baal library \cite{Atighehchian2020BAAL} and parallelized uncertainty calculations using the Ray package \cite{Ray2018}.

\begin{table}
\begin{center}
\caption{Training Hyperparameters for wav2vec 2.0 Model}
\label{table_hyperparams}
\begin{tabular}{|c|c|}
\hline
Hyperparameter & Value \\ 
\hline
Number of training epochs & 1000 \\ 
Learning rate & 4e-5 \\ 
Batch size & 32 \\ 
Gradient accumulation steps & 2 \\ 
Warm-up steps & 100 \\ 
Dropout rate & 0.1 \\
Loss function & CTC \\ 
Freeze encoder weights & Yes \\
\hline
\end{tabular}
\end{center}
\end{table}

\subsection{First Stage - Unsupervised Active Learning}
First, we will justify our choice of using x-vectors over i-vectors for clustering unlabeled data samples in the proposed AL approach. This decision is supported by the observation that x-vector representation, as opposed to i-vectors, provides a clearer separation of speech recordings from diverse data sources. To demonstrate this, we calculate Silhouette scores \cite{silhouette_score} for two groups of data: one comprising samples from the Common Voice dataset and the other from the whole LibriSpeech dataset.

The Silhouette score is a measure used to assess the separation between clusters. It calculates each point's average distance \( a(i) \) to the points in its own cluster and its distance \( b(i) \) to the points in the nearest cluster not containing it:
\begin{equation} \label{eq_silhouette_point}
s(i) = \frac{b(i) - a(i)}{\max(a(i), b(i))}.
\end{equation}
The overall Silhouette score is the mean of \( s(i) \) across all points in the dataset, which provides a clear measure of how distinct the clusters are. A higher Silhouette score indicates better-defined clusters that are well-separated from each other.

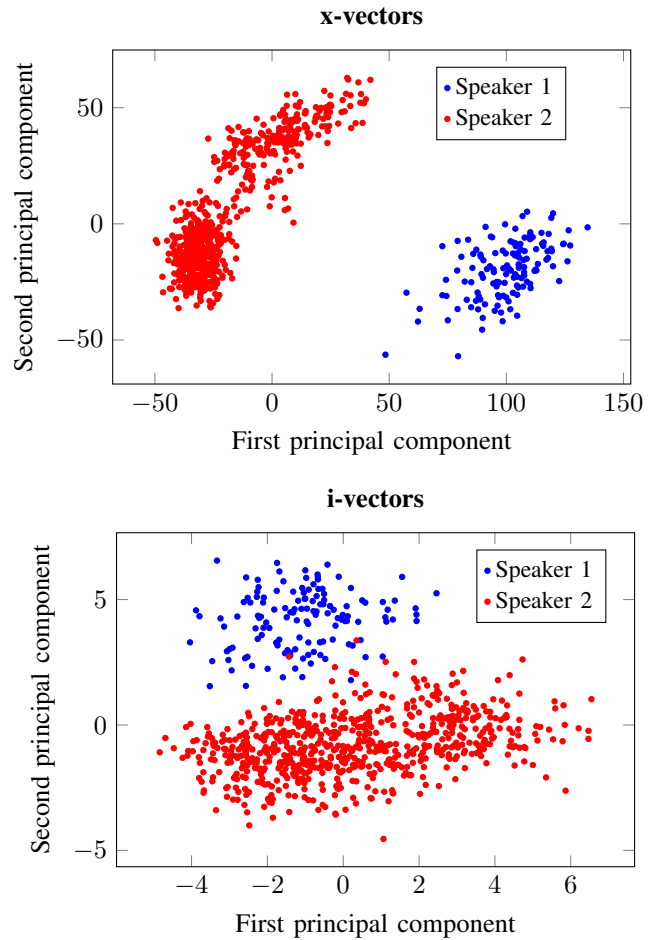
\begin{figure}[ht!]
\centering
\captionsetup[subfloat]{labelformat=empty}
\subfloat[]{
    \begin{tikzpicture}
    \begin{axis}[
        title={\textbf{x-vectors}},
        xlabel={First principal component},
        ylabel={Second principal component},
        legend entries={Speaker 1, Speaker 2},
        legend style={at={(0.75,0.95)},anchor=north, font=\small,},
        scatter/classes={0={mark=*,blue, mark size=1.0pt},1={mark=*,red, mark size=1.0pt}},
        height=6cm, 
        width=\axisdefaultwidth 
    ]
    \addplot[scatter,only marks,scatter src=explicit symbolic]
        table[col sep=comma,meta=cluster]{files/x_vector.csv};
    \end{axis}
    \end{tikzpicture}
}

\vspace{-0.7cm}

\subfloat[]{
    \begin{tikzpicture}
    \begin{axis}[
        title={\textbf{i-vectors}},
        xlabel={First principal component},
        ylabel={Second principal component},
        legend entries={Speaker 1, Speaker 2},
        legend style={at={(0.82,0.95)},anchor=north, font=\small,},
        scatter/classes={0={mark=*,blue, mark size=1.0pt},1={mark=*,red, mark size=1.0pt}},
        height=6cm, 
        width=\axisdefaultwidth 
    ]
    \addplot[scatter,only marks,scatter src=explicit symbolic]
        table[col sep=comma,meta=cluster]{files/i_vector.csv};
    \end{axis}
    \end{tikzpicture}
}

\vspace{-0.3cm}

\caption{The upper plot displays x-vectors and the lower plot shows i-vectors of speech recordings from two speakers, both reduced to two dimensions using PCA.}
\label{fig:x_and_i_vectors}
\end{figure}

The Silhouette scores, which measure the separation between the Common Voice and LibriSpeech datasets, were recorded at $0.0265$ for i-vectors and $0.0616$ for x-vectors, respectively. This considerable difference implies that x-vectors manage to better distinguish between the speech recordings from the two datasets. To further demonstrate this, we focus on two speakers, one from the Common Voice and the other from the LibriSpeech dataset. Fig. \ref{fig:x_and_i_vectors} presents the x-vector and i-vector representations of their speech recordings, reduced to two dimensions using principal component analysis (PCA). This visual comparison on a smaller example aligns with the conclusion of Silhouette analysis on the entire datasets, showing a more distinct separation for the x-vector representations. Thus, we opt for x-vector representations to facilitate clustering with DBSCAN, which forms a fundamental part of promoting diversity in our proposed approach. It is noteworthy that x-vectors were not specifically trained on these two datasets, indicating their robust generalization ability.

To demonstrate the effectiveness of the proposed unsupervised AL approach, we apply Algorithm \ref{first_stage_alg} to select $|S^0| = 643$ unlabeled samples (or approximately one hour of total duration), pair them with the corresponding labels, and use them as a training set for the wav2vec 2.0 ASR model. \green{We have compared the proposed unsupervised AL method against k-means clustering of i-vectors, k-means clustering of x-vectors, and DBSCAN clustering of i-vectors, and used random sampling as a baseline. Each of these methods also selects $643$ samples, and disproportionate cluster sampling is used for all methods, except for random sampling.} \green{Table \ref{table_first_stage} displays WER and CER on the primary test set for ASR models trained on datasets selected using these methods. The results indicate that DBSCAN outperforms k-means, whether using x-vectors or i-vectors, and that using x-vectors outperforms using i-vectors, regardless of the clustering method. Furthermore, our proposed approach, which uses DBSCAN over x-vectors, achieves the best performance, achieving lower WER and CER on the primary test set and demonstrating its effectiveness in selecting diverse and representative samples.}

\begin{table}
\caption{Comparison of the Proposed Unsupervised AL Approach With Random Sampling \green{and Alternative Methods (DBSCAN on i-vectors, k-means on x-vectors, and k-means on i-vectors)}}
\label{table_first_stage}
\begin{center}
\begin{tabular}{|c|c|c|}
\hline
Method & WER & CER \\
\hline
Random sampling & 0.2312 & 0.0678 \\
\textbf{Proposed approach (x-vectors DBSCAN)} & \textbf{0.2119} & \textbf{0.0613} \\
\green{i-vectors DBSCAN} & \multicolumn{1}{c|}{\green{0.2297}} & \multicolumn{1}{c|}{\green{0.0676}} \\
\green{x-vectors k-means} & \multicolumn{1}{c|}{\green{0.2254}} & \multicolumn{1}{c|}{\green{0.0661}} \\
\green{i-vectors k-means} & \multicolumn{1}{c|}{\green{0.2456}} & \multicolumn{1}{c|}{\green{0.0713}} \\
\hline
\end{tabular}
\end{center}
\end{table}

\subsection{Second Stage - Supervised Active Learning} \label{subsec-second-stage}

Next, we present the results of the second, supervised AL stage. In all experiments, we employed $T=20$ dropout models within a Bayesian committee, following the recommendation of at least ten MC dropout iterations as suggested by \cite{gal2016uncertainty}. After the initial unsupervised AL stage, we conducted $H=3$ supervised AL iterations, as described in Algorithm \ref{second_stage_alg}. During these iterations, we repeatedly trained wav2vec 2.0 ASR models\footnote{\blue{In this study, the ASR model is retrained from scratch in each AL iteration using all available labeled data. For scenarios with large amounts of labeled data, fine-tuning the model on newly collected data, starting from the model from the previous AL iteration, could be more practical. However, fine-tuning requires selecting additional parameters, such as a new learning rate, which could potentially lead to suboptimal results. To avoid this variability and ensure that we extract the maximum potential from the data selected in each iteration, we train the ASR model from scratch. Similarly, to fully leverage the selected data in each AL iteration, all models were trained for sufficiently long periods, up to 1000 epochs, with the best checkpoint selected after loss convergence, which consistently occurred between 102 and 235 epochs.}}, which identified and selected the most informative unlabeled samples based on the uncertainty metric. These samples were chosen from each x-vectors cluster to ensure diversity, resulting in a total of $|S^h|=643$ selected samples, which were then paired with a label and added to the training set.

We compared our supervised AL approach with \green{four} alternative methods:  
\begin{itemize}
    \item the signal-model committee approach (SMCA) proposed in [21],  
    \item an iterative random sampling strategy,  
    \item \green{an isolated first stage approach,}  
    \item \green{and an isolated second stage approach.}  
\end{itemize}
The SMCA includes a committee made up of the base model and one model modified using dropout, where uncertainty is measured by the character matching error rate (CMER) between the transcriptions produced by these two models. \green{In each AL iteration, SMCA selects the most uncertain samples. }\green{The isolated first stage approach uses only unsupervised AL for sample selection in each AL iteration, without proceeding to the second stage. In contrast, the isolated second stage approach replaces the first AL stage with random selection while keeping the supervised AL stage unchanged for the remaining iterations.} In each iteration of \green{all four} approaches, a total of \( |S^h|=643 \) samples are selected\green{, labeled, and added to the training set.} All wav2vec 2.0 ASR models used in our approach, SMCA, random sampling, \green{isolated first stage, and isolated second stage} were trained under the same conditions to ensure consistency in performance evaluation.

Before presenting the results across all supervised AL iterations, we assess the quality of uncertainty estimation of our Bayesian AL method by analyzing the Pearson correlations between the calculated uncertainties $U^h(x_i)$ for each primary test set sample and the WERs from the ASR model transcriptions relative to the ground-truth labels. These uncertainties are compared with those obtained using the SMCA method \cite{LMC-SMCA_2021_Committee1} and entropy-based uncertainties \cite{Active_learning_survey}. Entropy-based uncertainty is calculated by averaging the token entropy derived from the ASR model softmax layer distribution across the entire transcription. For these comparisons, we utilized the initial ASR model \(f_{\theta^{0}}\) trained on the labeled dataset selected during the first AL stage, $D_L^0$. The results are given in Table \ref{table_correlations}. It is evident that our approach achieves the highest Pearson correlation between the uncertainties and the actual WERs, indicating superior sample selection for enhancing model training. As expected, the entropy-based uncertainty method performed the least effectively, likely due to DNN overconfidence in predictions generated from the softmax layer \cite{Calibration_DNNs2017}, resulting in low entropy for both certain end uncertain samples. To display the relationship between the WERs and uncertainties calculated using our approach in a greater detail, these variables are presented in a scatter plot shown in Fig. \ref{fig_correlations}.

\begin{table}
\begin{center}
\caption{Pearson Correlations Between Test Set WERs and Different Uncertainty Measures}
\label{table_correlations}
\begin{tabular}{|c|c|}
\hline
Uncertainty & Pearson correlation \\ 
\hline
\textbf{Proposed method} & \textbf{0.5578} \\ 
SMCA & 0.4172 \\ 
Entropy & 0.3795 \\ 
\hline
\end{tabular}
\end{center}
\end{table}

\pgfplotstableread[col sep=comma]{files/scatter_results_wer.csv}\datawer
\pgfplotstableread[col sep=comma]{files/scatter_results.csv}\datauncertainty

\pgfplotstablecreatecol[copy column from table={\datauncertainty}{uncertainty}]{uncertainty}{\datawer}

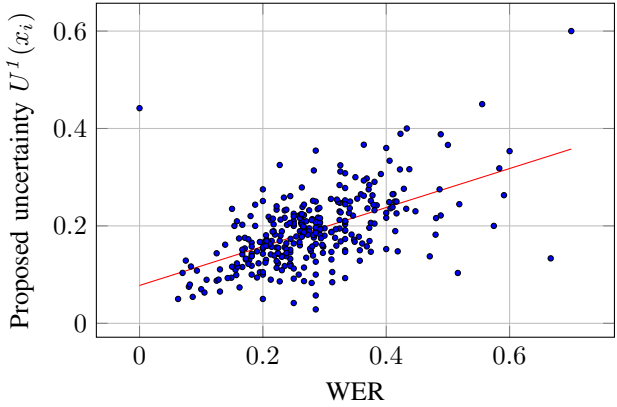
\begin{figure}[htbp]
\centering
\begin{tikzpicture}
\begin{axis}[
    xlabel={WER},
    ylabel={Proposed uncertainty $U^\mathit{1}(x_i)$},
    grid=major,
    height=6cm, 
    width=\axisdefaultwidth 
]
\addplot[only marks, mark size=1pt, fill=blue] table[x=wer, y=uncertainty] {\datawer};
\addplot [
        sharp plot,
        draw=red
    ] coordinates { 
        (0,0.077716649) (0.7,0.357772027592818)
    };
\end{axis}
\end{tikzpicture}
\caption{The relationship between WER and the proposed uncertainty on test set samples. The red line illustrates a linear fit to the provided relationship.}
\label{fig_correlations}
\end{figure}

\begin{figure}
\centering
\begin{tikzpicture}
\begin{axis}[
    xlabel={Iteration ($h$)},
    ylabel={Test Set WER (\%)},
    xmin=-0.2, xmax=3.2,
    ymin=7.5, ymax=24.5,
    xtick={0, 1, 2, 3},
    ytick={8, 10, 12, 14, 16, 18, 20, 22, 24, 26, 28},
    legend columns=2,  
    legend style={
        at={(0.5,1.3)},  
        anchor=north, 
        draw=none,  
        font=\small,  
        /tikz/every even column/.append style={column sep=5pt}  
    },
    ymajorgrids=true,
    grid style=dashed,
]

\addplot[
    color=blue,
    mark=square,
    ]
    coordinates {
    (0,21.19)(1,15.39)(2,12.86)(3,11.49)
    };
    \addlegendentry{Proposed approach}

\addplot[
    color=red,
    mark=triangle,
    ]
    coordinates {
    (0,23.12)(1,20.51)(2,18.14)(3,15.97)
    };
    \addlegendentry{SMCA}

\addplot[
    color=green,
    mark=o,
    ]
    coordinates {
    (0,23.12)(1,18.66)(2,15.82)(3,14.65)
    };
    \addlegendentry{Random sampling}

\addplot[
    color=orange,
    mark=star,
    ]
    coordinates {
    (0,21.19)(1,15.72)(2,14.02)(3,12.51)
    };
    \addlegendentry{\green{Isolated first stage}}

\addplot[
    color=magenta,
    mark=diamond,
    ]
    coordinates {
    (0,23.12)(1,16.65)(2,13.55)(3,11.89)
    };
    \addlegendentry{\green{Isolated second stage}}

\addplot[
    color=black,
    style=solid,
    line width=1.5pt,
]
    coordinates {
    (-0.2,8.85841363973)(3.2,8.85841363973)
    };
    \addlegendentry{The whole dataset}

\end{axis}
\end{tikzpicture}
\caption{Primary test set WER (\%) for trained ASR models over AL iterations, comparing the proposed approach, SMCA, random sampling,\green{ isolated first stage, isolated second stage,} and the whole dataset baseline. \blue{Each AL iteration adds approximately 1 hour of labeled training data, with the whole dataset totaling 17.31 hours.}}
\label{fig:results_plot}
\end{figure}
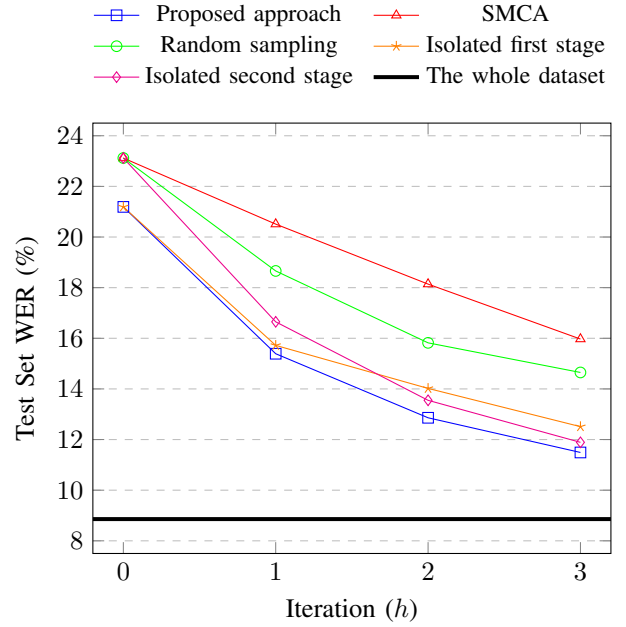

Fig. \ref{fig:results_plot} \green{compares the WER percentage} on the primary test set after training ASR models with additional newly labeled samples suggested by our method, SMCA, random sampling, \green{and the isolated first and second stages of our method} across successive AL iterations. Our approach consistently demonstrates lower test set WER in each AL iteration compared to competing approaches, indicating superior performance. \green{The results from the isolated first and second stages provide further insight into the contributions of each component of our method. Both isolated stages outperform SMCA and random sampling across all iterations, confirming the benefits of diversity-based and uncertainty-based selection compared to standard baselines. The isolated first stage achieves lower WER in the first two iterations because it begins with a more diverse and representative dataset, whereas the isolated second stage starts with a randomly sampled initial dataset. However, in the last two iterations, the isolated second stage surpasses the first stage, suggesting that uncertainty-based selection becomes increasingly effective as the model improves. The proposed two-stage approach consistently outperforms both isolated stages, demonstrating that the combination of diversity-driven selection and uncertainty-based refinement is crucial for maximizing AL performance.}

\green{SMCA is outperformed by the iterative random sampling approach due to two key factors. One of the reasons is that its uncertainty measure correlates less with the actual WER compared to our proposed method, as shown in Table IV, which results in a slight reduction in its effectiveness when selecting informative samples. In addition, and more importantly, while the test set is designed to evaluate models on diverse data, SMCA in our case tends to select the most uncertain samples, which likely come from a small number of clusters. This leads to a less diverse training set for SMCA-based ASR models, further limiting their performance. In contrast, random sampling selects data with greater diversity, leading to a more balanced training set. This confirms the competitiveness of random sampling under certain conditions, as discussed in \cite{Munjal2022Cvpr_random_sampling}.} Given that the primary test set includes underrepresented groups of speakers from the training set, the diversity-driven approach of our method, especially our use of disproportionate cluster sampling that favors smaller clusters, is particularly crucial. Our method combines the precision of uncertainty-based selection with the advantages of diversity, which together account for its enhanced performance.

Additionally, the horizontal black line in Fig. \ref{fig:results_plot} represents the WER when the model is trained with the whole dataset\blue{, which has a total duration of 17.31 hours.} This serves as a benchmark compared to the much smaller datasets selected through AL iterations, indicating the performance achievable with the current hyperparameters and available training data. The proximity of our AL method's performance to this benchmark underscores the efficiency of using AL strategies: achieving competitive performance while training the ASR model with only $19.98\%$ of the whole training dataset. This comparison emphasizes the benefits of our AL approach, which leverages a smaller, more targeted subset of the whole dataset, thereby reducing labeling effort, training time and resource consumption while maintaining high accuracy.

\begin{figure}
\centering
\begin{tikzpicture}
\begin{axis}[
    xlabel={Iteration ($h$)},
    ylabel={Uncertainty},
    ymin=-0.05, ymax=0.55,
    xtick={1, 2, 3},
    ytick={0, 0.5, 1.0, 1.5, 2.0},
    ylabel near ticks,
    width=\linewidth, 
    height=6cm, 
    grid=major,
    boxplot/draw direction=y,
    boxplot={
        box extend=0.3,
    },
    every axis plot/.append style={thick}
]

\addplot+[
    boxplot prepared={
        median=0.116666666666667,
        lower quartile=0.05,
        upper quartile=0.207142857142857,
        upper whisker=0.441666666666667,
        lower whisker=0.0
    },
    fill=blue!30,
    draw=black
] coordinates {};

\addplot+[
    boxplot prepared={
        median=0.0833333333333333,
        lower quartile=0.025,
        upper quartile=0.1555555555555556,
        upper whisker=0.3500000000000001,
        lower whisker=0.0
    },
    fill=blue!30,
    draw=black
] coordinates {};

\addplot+[
    boxplot prepared={
        median=0.059375,
        lower quartile=0.0125,
        upper quartile=0.1214285714285714,
        upper whisker=0.2846153846153846,
        lower whisker=0.0    
    },
    fill=blue!30,
    draw=black
] coordinates {};

\end{axis}
\end{tikzpicture}
\caption{The distribution of uncertainties calculated using the proposed approach for all unlabeled samples in each AL iteration.}
\label{fig:whiskers}
\end{figure}
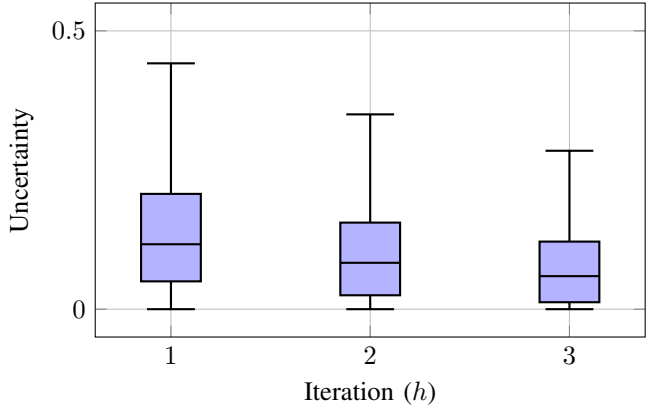

To better understand how uncertainty is reduced over time using our approach, we observe the uncertainties calculated by the Bayesian committee after each AL iteration. This allows us to understand better how the proposed data selection refines ASR model confidence progressively. Fig. \ref{fig:whiskers} presents whisker plots that demonstrate the uncertainty distributions of all unlabeled samples after each AL iteration. The consistent decrease in uncertainty across the iterations confirms that our approach effectively enhances model certainty over time. Notably, the diminishing length of the upper whisker over successive iterations indicates a significant reduction in higher uncertainty samples.

\subsection{Results on OOD Test Set}

\begin{figure}
\centering
\begin{tikzpicture}
\begin{axis}[
    xlabel={Iteration ($h$)},
    ylabel={Test Set WER (\%)},
    xmin=-0.2, xmax=3.2,
    ymin=23.5, ymax=38.5,
    xtick={0, 1, 2, 3},
    ytick={24, 26, 28, 30, 32, 34, 36, 38},
    legend style={
        at={(0.44,0.65)},
        anchor=south west, 
        font=\small,
    },
    ymajorgrids=true,
    grid style=dashed,
]

\addplot[
    color=blue,
    mark=square,
    ]
    coordinates {
    (0,35.812454938716654)(1,31.27928983417448)(2,28.746845)(3,27.81407714491709)
    };
    \addlegendentry{Proposed approach}

\addplot[
    color=red,
    mark=triangle,
    ]
    coordinates {
    (0,36.96151766402307)(1,33.09525955299207)(2,31.308579)(3,29.5421773612112)
    };
    \addlegendentry{SMCA}

\addplot[
    color=green,
    mark=o,
    ]
    coordinates {
    (0,36.96151766402307)(1,32.70322638788753)(2,30.4253785147801)(3,29.55569574621485)
    };
    \addlegendentry{Random sampling}

\addplot[
    color=black,
    style=solid,
    line width=1.5pt,
]
    coordinates {
    (-0.2,24.64176279740447)(3.2,24.64176279740447)
    };
    \addlegendentry{The whole dataset}

\end{axis}
\end{tikzpicture}
\caption{OOD test set WER (\%) for ASR models previously trained in each AL iteration, comparing the proposed approach, SMCA, random sampling, and the whole \blue{17.31-hour} dataset baseline.}
\label{fig:ood}
\end{figure}
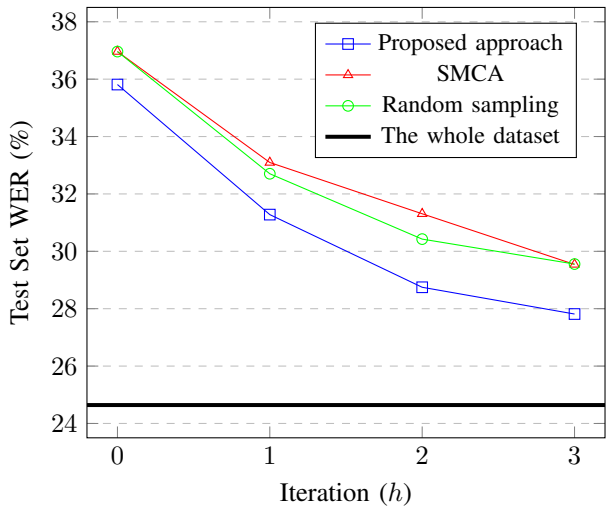

To further evaluate our approach, we tested its performance on the heterogeneous VoxPopuli test set \cite{wang2021voxpopuli}, where the domain-specific content and language induce a notable shift in data distribution. The evaluation outcome is presented in Fig. \ref{fig:ood}, which displays the OOD test set WER for each ASR model trained in Section \ref{subsec-second-stage}, including models trained with data iteratively selected by the proposed approach, SMCA, and random sampling, as well as a model trained on the whole \blue{17.31-hour} training set. Expectedly, WERs in this test scenario are higher compared to those in Section \ref{subsec-second-stage} due to the significant distribution shift in the test data and the relatively brief training duration paired with relatively small training set sizes. The horizontal black line denotes the model trained on the whole dataset, serving as a benchmark compared to the much smaller datasets selected through AL iterations; its high WER confirms the distribution shift in the OOD dataset. The results indicate that while the performances of the proposed approach, SMCA, and random sampling are relatively close, our approach demonstrates superior results on the OOD test set. The increasing difference in OOD test set WER can be attributed to the diverse training set samples suggested by our approach, with the advantage becoming more pronounced as we further add newly labeled samples.

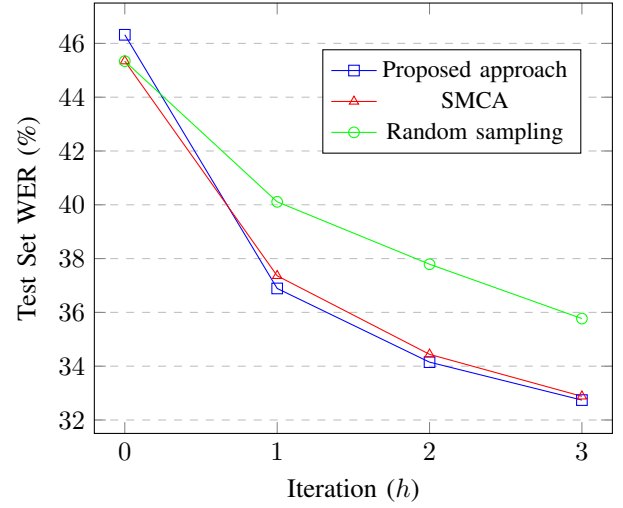
\begin{figure}
\centering
\begin{tikzpicture}
\begin{axis}[
    xlabel={Iteration ($h$)},
    ylabel={Test Set WER (\%)},
    xmin=-0.2, xmax=3.2,
    ymin=31.5, ymax=47.5,
    xtick={0, 1, 2, 3},
    ytick={32, 34, 36, 38, 40, 42, 44, 46},
    legend style={
        at={(0.44,0.65)},
        anchor=south west, 
        font=\small,
    },
    ymajorgrids=true,
    grid style=dashed,
]

\addplot[
    color=blue,
    mark=square,
    ]
    coordinates {
    (0,46.32)(1,36.89)(2,34.15)(3,32.74)
    };
    \addlegendentry{Proposed approach}

\addplot[
    color=red,
    mark=triangle,
    ]
    coordinates {
    (0,45.34)(1,37.36)(2,34.44)(3,32.88)
    };
    \addlegendentry{SMCA}

\addplot[
    color=green,
    mark=o,
    ]
    coordinates {
    (0,45.34)(1,40.11)(2,37.79)(3,35.77)
    };
    \addlegendentry{Random sampling}

\end{axis}
\end{tikzpicture}
\caption{\green{WER (\%) on the Common Voice test set for ASR models trained in each AL iteration, comparing the proposed approach, SMCA, and random sampling.}}
\label{fig:commonvoice}
\end{figure}

\green{
\subsection{Results on a Standard ASR Evaluation Benchmark}

To evaluate whether the proposed approach maintains competitive performance on a widely used ASR benchmark, we tested it on the standard Common Voice train-test split. In this setup, the Common Voice train split containing 12135 samples serves as the unlabeled dataset for AL, from which training samples are iteratively selected, while ASR models are evaluated on the Common Voice test split containing 7015 samples. This evaluation ensures that the improvements observed on the primary and OOD test sets do not come at the cost of overall ASR performance on a standard dataset. As in the primary test set experiment, we first perform the unsupervised AL stage, followed by three iterations of the supervised AL stage, selecting 643 samples per iteration from the unlabeled dataset and training ASR models accordingly.

The results on the Common Voice test set are shown in Fig. \ref{fig:commonvoice}, comparing the proposed approach to SMCA and random sampling. As in previous evaluations, SMCA was initialized with the ASR model trained on the initial random sampling selection. In the first AL iteration, SMCA and random sampling slightly outperform the proposed approach. This can occur when the unlabeled and test data are largely drawn from the dominant speaker group, creating a distributional similarity between them. Unlike random sampling, the proposed first AL stage can detect small clusters within the unlabeled dataset and enforce sample selection from them, even if similar samples do not exist in the test set. As a result, fewer samples from the dominant speaker group are selected in the first AL iteration, causing deteriorated test set performance. This aligns with the fact that the standard Common Voice test set is heterogeneous and not specifically designed to evaluate performance on underrepresented speaker groups.

However, in later AL iterations, the proposed method achieves the lowest WER, particularly compared to random sampling, demonstrating the benefits of the uncertainty-based second AL stage. This is further supported by the competitive performance of the purely uncertainty-based SMCA, which closely follows the proposed approach on this benchmark, in contrast to its weaker performance on the primary test set, where the focus is on evaluating underrepresented speaker groups.
}

\section{Conclusion}

This study developed and implemented a two-stage AL pipeline for ASR, aimed at enhancing model training performance while minimizing labeling effort. The first AL stage utilizes unsupervised AL with x-vectors clustering to select a diverse initial subset of unlabeled speech samples. Once labeled, these samples form a high-quality and diverse dataset that establishes a robust foundation for the subsequent supervised AL stage. Our results show that this initial selection significantly improves the test set performance of the initial ASR model compared to the random sampling approach, indicating effective dataset curation.

In the second AL stage, we adopt a supervised AL strategy that features a batch AL method customized for ASR. This stage promotes further diversity through x-vectors clustering and also incorporates a Bayesian AL method with MC dropout, specifically adapted for uncertainty estimation in ASR. When evaluated on a test set representing underrepresented speaker groups, our supervised AL method consistently outperforms competing methods, underscoring the benefits of promoting diversity through x-vectors clustering. Additionally, the proposed approach outperforms competing methods on the heterogeneous OOD test set,\green{ while on the standard ASR benchmark, its first stage performs slightly worse initially, but the second stage achieves the best results in later iterations.}

\section*{Acknowledgment}
The computational resources required for the experiments in this paper were provided by the Serbian National Platform for Artificial Intelligence, an initiative of the Office for Information Technologies and e-Government.


\bibliographystyle{IEEEtran}
\bibliography{refs}

\begin{thebibliography}{10}
\providecommand{\url}[1]{#1}
\csname url@samestyle\endcsname
\providecommand{\newblock}{\relax}
\providecommand{\bibinfo}[2]{#2}
\providecommand{\BIBentrySTDinterwordspacing}{\spaceskip=0pt\relax}
\providecommand{\BIBentryALTinterwordstretchfactor}{4}
\providecommand{\BIBentryALTinterwordspacing}{\spaceskip=\fontdimen2\font plus
\BIBentryALTinterwordstretchfactor\fontdimen3\font minus \fontdimen4\font\relax}
\providecommand{\BIBforeignlanguage}[2]{{%
\expandafter\ifx\csname l@#1\endcsname\relax
\typeout{** WARNING: IEEEtran.bst: No hyphenation pattern has been}%
\typeout{** loaded for the language `#1'. Using the pattern for}%
\typeout{** the default language instead.}%
\else
\language=\csname l@#1\endcsname
\fi
#2}}
\providecommand{\BIBdecl}{\relax}
\BIBdecl

\bibitem{McMullin2023TranscriptionTime}
C.~McMullin, ``{Transcription and qualitative methods: Implications for third sector research},'' \emph{Voluntas: Int. J. Volunt. Nonproﬁt Organ.}, vol.~34, no.~1, pp. 140--153, 2023.

\bibitem{NEURIPS2023_DataCentricAI}
M.~Mazumder \emph{et~al.}, ``{DataPerf: Benchmarks for Data-Centric AI Development},'' in \emph{Proc. 36th Int. Conf. Neural Inf. Process. Syst.}, 2023, pp. 5320--5347.

\bibitem{DataCentricAI}
S.~Kumar, S.~Datta, V.~Singh, S.~K. Singh, and R.~Sharma, ``{Opportunities and Challenges in Data-Centric AI},'' \emph{IEEE Access}, vol.~12, pp. 33\,173--33\,189, 2024.

\bibitem{DataSelection2007}
Y.~Wu, R.~Zhang, and A.~Rudnicky, ``{Data selection for speech recognition},'' in \emph{2007 IEEE Workshop Autom. Speech Recognit. Understand.}, pp. 562--565.

\bibitem{lee-etal-2022-deduplicating}
K.~Lee, D.~Ippolito, A.~Nystrom, C.~Zhang, D.~Eck, C.~Callison-Burch, and N.~Carlini, ``{Deduplicating Training Data Makes Language Models Better},'' in \emph{Proc. 60th Ann. Meeting Assoc. Comput. Linguist.}, 2022, pp. 8424--8445.

\bibitem{DataManagementForML2023}
C.~Chai, J.~Wang, Y.~Luo, Z.~Niu, and G.~Li, ``{Data Management for Machine Learning: A Survey},'' \emph{IEEE Trans. Knowl. Data Eng.}, vol.~35, no.~5, pp. 4646--4667, 2023.

\bibitem{2021SurveyDeepActiveLearning}
P.~Ren, Y.~Xiao, X.~Chang, P.-Y. Huang, Z.~Li, B.~B. Gupta, X.~Chen, and X.~Wang, ``{A Survey of Deep Active Learning},'' \emph{ACM Comput. Surv.}, vol.~54, no.~9, 2021.

\bibitem{AL_ASR_2005}
G.~Riccardi and D.~Hakkani-Tur, ``{Active learning: theory and applications to automatic speech recognition},'' \emph{IEEE Trans. Speech Audio Process.}, vol.~13, no.~4, pp. 504--511, 2005.

\bibitem{YU2010GlobalEntropy}
D.~Yu, B.~Varadarajan, L.~Deng, and A.~Acero, ``{Active learning and semi-supervised learning for speech recognition: A unified framework using the global entropy reduction maximization criterion},'' \emph{Comput. Speech Lang.}, vol.~24, pp. 433--444, 2010.

\bibitem{endToEndASRSurvey2024}
R.~Prabhavalkar, T.~Hori, T.~N. Sainath, R.~Schlüter, and S.~Watanabe, ``End-to-end speech recognition: A survey,'' \emph{IEEE/ACM Trans. Audio Speech Lang. Process.}, vol.~32, pp. 325--351, 2024.

\bibitem{Bang2020BoostingAL}
J.~Bang, H.~Kim, Y.~Yoo, and J.-W. Ha, ``Boosting active learning for speech recognition with noisy pseudo-labeled samples,'' vol. arXiv: 2006.11021, 2020.

\bibitem{drugman16_interspeech}
T.~Drugman, J.~Pylkkönen, and R.~Kneser, ``{Active and Semi-Supervised Learning in ASR: Benefits on the Acoustic and Language Models},'' in \emph{Proc. Interspeech}, 2016, pp. 2318--2322.

\bibitem{Indian2018AL_ASR}
M.~Chellapriyadharshini, A.~Toffy, S.~R.~K. M., and V.~Ramasubramanian, ``{Semi-supervised and Active-learning Scenarios: Efficient Acoustic Model Refinement for a Low Resource Indian Language},'' in \emph{Proc. Interspeech}, 2018, pp. 1--5.

\bibitem{Calibration_DNNs2017}
C.~Guo, G.~Pleiss, Y.~Sun, and K.~Q. Weinberger, ``{On calibration of modern neural networks},'' in \emph{Proc. 34th Int. Conf. Mach. Learn.}, 2017, p. 1321–1330.

\bibitem{jiaji2016Gradients1}
J.~Huang, R.~Child, V.~Rao, H.~Liu, S.~Satheesh, and A.~Coates, ``{Active learning for speech recognition: The power of gradients},'' in \emph{Proc. 30th Int. Conf. Neural Inf. Process. Syst.}, 2016, pp. 1--5.

\bibitem{Yuan2019Gradients2}
Y.~Yuan, S.-W. Chung, and H.-G. Kang, ``{Gradient-based Active Learning Query Strategy for End-to-end Speech Recognition},'' in \emph{Proc. 2019 IEEE Int. Conf. Acoust. Speech Signal Process.}, pp. 2832--2836.

\bibitem{2021LossPrediction}
J.~Luo, J.~Wang, N.~Cheng, and J.~Xiao, ``{Loss Prediction: End-to-End Active Learning Approach For Speech Recognition},'' in \emph{2021 Int. Joint Conf. Neural Netw.}, pp. 1--7.

\bibitem{i_vectors}
N.~Dehak, P.~J. Kenny, R.~Dehak, P.~Dumouchel, and P.~Ouellet, ``{Front-End Factor Analysis for Speaker Verification},'' \emph{IEEE Trans. Audio Speech Lang. Process.}, vol.~19, no.~4, pp. 788--798, 2011.

\bibitem{malhotra19_i_vectors}
K.~Malhotra, S.~Bansal, and S.~Ganapathy, ``{Active Learning Methods for Low Resource End-to-End Speech Recognition},'' in \emph{Proc. Interspeech}, 2019, pp. 2215--2219.

\bibitem{LF_MMI_interspeech2018}
Y.~Long, H.~Ye, Y.~Li, and J.~Liang, ``{Active Learning for LF-MMI Trained Neural Networks in ASR},'' in \emph{Proc. Interspeech}, 2018, pp. 2898--2902.

\bibitem{LMC-SMCA_2021_Committee1}
{Sun, Xiusong and Wang, Bo and Liu, Shaohan and Lu, Tingxiang and Shan, Xin and Yang, Qun}, ``Lmc-smca: A new active learning method in asr,'' \emph{IEEE Access}, vol.~9, pp. 37\,011--37\,021, 2021.

\bibitem{2019CommiteeDropout}
J.~Fu and K.~Ru, ``{A Dropout-Based Single Model Committee Approach for Active Learning in ASR},'' in \emph{2019 IEEE Workshop Autom. Speech Recognit. Understand.}, pp. 16--22.

\bibitem{unsupervised_AL_ASR}
A.~R. Syed, A.~Rosenberg, and E.~Kislal, ``{Supervised and unsupervised active learning for automatic speech recognition of low-resource languages},'' in \emph{Proc. 2016 IEEE Int. Conf. Acoust. Speech Signal Process.}, pp. 5320--5324.

\bibitem{unsupervisedAL_interspeech2023}
Z.~Zheng, Z.~Ma, Y.~Wang, and X.~Chen, ``{Unsupervised Active Learning: Optimizing Labeling Cost-Effectiveness for Automatic Speech Recognition},'' in \emph{Proc. Interspeech}, 2023, pp. 3307--3311.

\bibitem{NEURIPS2021_UnsupervisedASR}
A.~Baevski, W.-N. Hsu, A.~Conneau, and M.~Auli, ``{Unsupervised Speech Recognition},'' in \emph{Proc. 34th Int. Conf. Neural Inf. Process. Syst.}, 2021, pp. 27\,826--27\,839.

\bibitem{hubert2021}
W.-N. Hsu, B.~Bolte, Y.-H.~H. Tsai, K.~Lakhotia, R.~Salakhutdinov, and A.~Mohamed, ``Hubert: Self-supervised speech representation learning by masked prediction of hidden units,'' \emph{IEEE/ACM Trans. Audio Speech Lang. Process.}, vol.~29, pp. 3451--3460, 2021.

\bibitem{selfSupervisedForASR2024}
H.~Zhu, G.~Cheng, J.~Wang, W.~Hou, P.~Zhang, and Y.~Yan, ``Boosting cross-domain speech recognition with self-supervision,'' \emph{IEEE/ACM Trans. Audio Speech Lang. Process.}, vol.~32, pp. 471--485, 2024.

\bibitem{wav2vec2}
A.~Baevski, H.~Zhou, A.~Mohamed, and M.~Auli, ``{wav2vec 2.0: a framework for self-supervised learning of speech representations},'' in \emph{Proc. 34th Int. Conf. Neural Inf. Process. Syst.}, 2021, pp. 12\,449--12\,460.

\bibitem{x_vectors}
D.~Snyder, D.~Garcia-Romero, G.~Sell, D.~Povey, and S.~Khudanpur, ``{X-Vectors: Robust DNN Embeddings for Speaker Recognition},'' in \emph{Proc. 2018 IEEE Int. Conf. Acoust. Speech Signal Process.}, pp. 5329--5333.

\bibitem{BayesianNNbook}
R.~M. Neal, \emph{{Bayesian Learning for Neural Networks}}.\hskip 1em plus 0.5em minus 0.4em\relax Berlin, Heidelberg: Springer-Verlag, 1996.

\bibitem{pmlr-MC_dropout}
Y.~Gal and Z.~Ghahramani, ``{Dropout as a Bayesian Approximation: Representing Model Uncertainty in Deep Learning},'' in \emph{Proc. 33rd Int. Conf. Mach. Learn.}, 2016, pp. 1050--1059.

\bibitem{GIDIOTIS2024BayesianActiveSummar}
A.~Gidiotis and G.~Tsoumakas, ``{Bayesian active summarization},'' \emph{Comput. Speech Lang.}, vol.~83, p. 101553, 2024.

\bibitem{Graves2006CTC}
A.~Graves, S.~Fern\'{a}ndez, F.~Gomez, and J.~Schmidhuber, ``{Connectionist temporal classification: labelling unsegmented sequence data with recurrent neural networks},'' in \emph{Proc. 23th Int. Conf. Mach. Learn.}, 2006, p. 369–376.

\bibitem{Active_learning_survey}
C.~Aggarwal, X.~Kong, Q.~Gu, J.~Han, and P.~Yu, \emph{\BIBforeignlanguage{English (US)}{{Active learning: A survey}}}.\hskip 1em plus 0.5em minus 0.4em\relax CRC Press, Jan. 2014.

\bibitem{DAGAN1995Committee}
I.~Dagan and S.~P. Engelson, ``{Committee-Based Sampling For Training Probabilistic Classifiers},'' in \emph{Machine Learning Proceedings 1995}, A.~Prieditis and S.~Russell, Eds.\hskip 1em plus 0.5em minus 0.4em\relax San Francisco (CA): Morgan Kaufmann, 1995, pp. 150--157.

\bibitem{Houlsby2011BayesianAL}
N.~Houlsby, F.~Huszar, Z.~Ghahramani, and M.~Lengyel, ``{Bayesian Active Learning for Classification and Preference Learning},'' vol. arXiv: 1112.5745, 2011.

\bibitem{gal2016uncertainty}
Y.~Gal, ``{Uncertainty in deep learning},'' Ph.D. dissertation, University of Cambridge, 2016.

\bibitem{NEURIPS2021_BatchAL_At_scale}
G.~Citovsky, G.~DeSalvo, C.~Gentile, L.~Karydas, A.~Rajagopalan, A.~Rostamizadeh, and S.~Kumar, ``{Batch Active Learning at Scale},'' in \emph{Proc. 34th Int. Conf. Neural Inf. Process. Syst.}, M.~Ranzato, A.~Beygelzimer, Y.~Dauphin, P.~Liang, and J.~W. Vaughan, Eds., 2021, pp. 11\,933--11\,944.

\bibitem{DBSCAN2017}
E.~Schubert, J.~Sander, M.~Ester, H.~P. Kriegel, and X.~Xu, ``{DBSCAN Revisited, Revisited: Why and How You Should (Still) Use DBSCAN},'' \emph{ACM Trans. Database Syst.}, vol.~42, no.~3, jul 2017.

\bibitem{NEURIPS2019_BatchBALD}
A.~Kirsch, J.~van Amersfoort, and Y.~Gal, ``{BatchBALD: Efficient and Diverse Batch Acquisition for Deep Bayesian Active Learning},'' in \emph{Proc. 32nd Int. Conf. Neural Inf. Process. Syst.}, H.~Wallach, H.~Larochelle, A.~Beygelzimer, F.~d\textquotesingle Alch\'{e}-Buc, E.~Fox, and R.~Garnett, Eds., 2019, p. 7024–7035.

\bibitem{Atighehchian2020BAAL}
P.~Atighehchian, F.~Branchaud-Charron, and A.~Lacoste, ``{Bayesian active learning for production, a systematic study and a reusable library},'' vol. arXiv: 2006.09916, 2020.

\bibitem{common-voice-2020-common}
R.~Ardila, M.~Branson, K.~Davis, M.~Kohler, J.~Meyer, M.~Henretty, R.~Morais, L.~Saunders, F.~Tyers, and G.~Weber, ``\BIBforeignlanguage{English}{{Common Voice: A Massively-Multilingual Speech Corpus}},'' in \emph{\BIBforeignlanguage{English}{Proc. 12th Lang. Resour. Eval. Conf.}}, Marseille, France, May 2020, pp. 4218--4222.

\bibitem{LibriSpeech2015}
V.~Panayotov, G.~Chen, D.~Povey, and S.~Khudanpur, ``{Librispeech: An ASR corpus based on public domain audio books},'' in \emph{Proc. 2015 IEEE Int. Conf. Acoust. Speech Signal Process.}, pp. 5206--5210.

\bibitem{Munjal2022Cvpr_random_sampling}
P.~Munjal, N.~Hayat, M.~Hayat, J.~Sourati, and S.~Khan, ``{Towards Robust and Reproducible Active Learning using Neural Networks},'' in \emph{2022 Proc. IEEE Conf. Comput. Vis. Pattern Recognit.}\hskip 1em plus 0.5em minus 0.4em\relax Los Alamitos, CA, USA: IEEE Computer Society, jun, pp. 223--232.

\bibitem{DiChristofano2024}
A.~DiChristofano, H.~Shuster, S.~Chandra, and N.~Patwari, ``Performance disparities between accents in automatic speech recognition,'' in \emph{Proc. AAAI Conf. Artif. Intell.}, vol.~37, no.~13, 2024, pp. 16\,200--16\,201.

\bibitem{tadimeti-2022}
D.~Tadimeti, K.~Georgila, and D.~Traum, ``Evaluation of off-the-shelf speech recognizers on different accents in a dialogue domain,'' in \emph{Proc. 13th Lang. Resour. Eval. Conf.}, Marseille, France, 2022, pp. 6001--6008.

\bibitem{Graham2024}
C.~Graham and N.~Roll, ``Evaluating openai's whisper asr: Performance analysis across diverse accents and speaker traits,'' \emph{JASA Express Lett.}, vol.~4, no.~2, p. 025206, 2024.

\bibitem{Allison2020}
A.~Koenecke, A.~Nam, E.~Lake, J.~Nudell, M.~Quartey, Z.~Mengesha, C.~Toups, J.~R. Rickford, D.~Jurafsky, and S.~Goel, ``Racial disparities in automated speech recognition,'' \emph{Proc. Natl. Acad. Sci. U.S.A.}, vol. 117, no.~14, pp. 7684--7689, 2020.

\bibitem{wang2021voxpopuli}
C.~Wang, M.~Riviere, A.~Lee, A.~Wu, C.~Talnikar, D.~Haziza, M.~Williamson, J.~Pino, and E.~Dupoux, ``{VoxPopuli: A Large-Scale Multilingual Speech Corpus for Representation Learning, Semi-Supervised Learning and Interpretation},'' in \emph{Proc. 59th Ann. Meeting Assoc. Comput. Linguist.}, 2021.

\bibitem{Babu2021XLSRSC}
A.~Babu, C.~Wang, A.~Tjandra, K.~Lakhotia, Q.~Xu, N.~Goyal, K.~Singh, P.~{von Platen}, Y.~Saraf, J.~Pino, A.~Baevski, A.~Conneau, and M.~Auli, ``{XLS-R: Self-supervised Cross-lingual Speech Representation Learning at Scale},'' in \emph{Proc. Interspeech}, 2022, pp. 2278--2282.

\bibitem{Ray2018}
P.~Moritz, R.~Nishihara, S.~Wang, A.~Tumanov, R.~Liaw, E.~Liang, M.~Elibol, Z.~Yang, W.~Paul, M.~I. Jordan, and I.~Stoica, ``{Ray: a distributed framework for emerging AI applications},'' in \emph{Proc. 13th USENIX Conf. Operat. Syst. Des. Implement.}, 2018, p. 561–577.

\bibitem{silhouette_score}
P.~J. Rousseeuw, ``{Silhouettes: A graphical aid to the interpretation and validation of cluster analysis},'' \emph{J. Comput. Appl. Math.}, vol.~20, pp. 53--65, 1987.

\end{thebibliography}

\vfill

\end{document}